\documentclass[a4paper,fleqn]{cas-dc}

\usepackage[numbers]{natbib}

\def\tsc#1{\csdef{#1}{\textsc{\lowercase{#1}}\xspace}}
\tsc{WGM}
\tsc{QE}

\begin{document}
\let\WriteBookmarks\relax
\def\floatpagepagefraction{1}
\def\textpagefraction{.001}

\shorttitle{Microwave scattering by rough polyhedral particles on a surface}    

\shortauthors{Virkki and Yurkin}  

\title[mode = title]{Microwave scattering by rough polyhedral particles on a surface}  



%

\author[1,2]{Anne K. Virkki}[type=editor, orcid=0000-0002-4129-5381]

\cormark[1]


\ead{anne.virkki@helsinki.fi}


\credit{Conceptualization, Project administration, Funding acquisition, Methodology, Software, Investigation, Formal analysis, Visualization, Data curation, Writing -- original draft, review, and editing, Validation}

\affiliation[1]{organization={University of Helsinki},
            addressline={Gustaf H\"{a}llstr\"{o}min katu 2}, 
            city={Helsinki},
            postcode={00560}, 
            country={Finland}}

\affiliation[2]{organization={Finnish Geospatial Research Institute},
            addressline={Vuorimiehentie 5}, 
            city={Espoo},
            postcode={02150}, 
            country={Finland}}
            
\author[3]{Maxim A. Yurkin}[orcid=0000-0002-3524-0093]


\ead{yurkin@gmail.com}


\credit{Methodology, Investigation, Formal analysis, Writing -- original draft, review, and editing, Validation}

\affiliation[3]{organization={Universit\'{e} Rouen Normandie, INSA Rouen Normandie, CNRS, CORIA UMR 6614},
	city={Rouen},
	postcode={F-76000}, 
	country={France}}

\cortext[1]{Corresponding author}



\begin{abstract}
The electromagnetic (EM) scattering by non-symmetric wavelength-scale particles on a planar surface has numerous applications in the remote sensing of planetary bodies, both in planetary and geo-sciences. We conduct numerical simulations of EM scattering by rough polyhedral particles (with 12 or 20 faces) using the discrete-dipole approximation and contrast the results to that of spheres. The particles have permittivities corresponding to common minerals in the microwave regime ($\epsilon_r=4.7 + 0.016$i and $7.8 + 0.09$i), and a size-frequency distribution (SFD) consistent with the observed scattering properties (power-law distribution of size parameters between 0.5 and 8 with an index from $-2.5$ to $-3.5$). The assumed substrate permittivity $2.4 + 0.012$i corresponds to a powdered regolith. We present what roles the particle roundness, permittivity, and SFD for a realistic range of parameters play in the EM scattering properties as a function of incidence angle with a focus on backscattering in microwave-remote-sensing applications. The particle roundness and SFD have a clearly observable effect on the polarimetric properties, while the role of permittivity is relatively minor (in the studied range). Among various backscattering observables, the circular polarization ratio is the least sensitive to the decrease of the upper boundary (down to a size parameter of 3) and the index of the SFD.
\end{abstract}


\begin{highlights}
\item Systematic light-scattering simulations for polyhedral particles on a surface.
\item Microwave (radar) observations of rocky surfaces are discussed as an application.
\item The particle roundness and size distribution have a clearly observable effect.
\item The role of permittivity is relatively minor among typical minerals.
\end{highlights}



\maketitle

\section{Introduction}\label{sec:intro}

The surfaces of atmosphereless planetary objects are typically composed of regolith, a loose fine-grained rocky substance with unspecified mineralogical composition. The particle sizes extend from micrometers to meters depending on the object: the Moon and other large objects tend to have dominantly micrometer-scale regolith with sparse meter-scale boulders, whereas on smaller asteroids of up to a few hundred meters, the visible surfaces can be dominantly composed of larger, millimeter-to-centimeter scale grains and meter-scale boulders are more common, as for example the recent Hayabusa2 mission to asteroid (162173) Ryugu \cite{watanabe2019} and the OSIRIS-REx mission to (101955) Bennu \cite{lauretta2019} revealed. Although a handful of asteroids have been visited by spacecraft, typically the only source of information about asteroids is Earth-based observations. Therefore, it is crucial to understand the scattering properties of planetary materials to find their diverse physical properties through inverse modeling. 

In this work, we focus on remote sensing of planetary surfaces using microwaves. In the observations of rocky surfaces, there is often a wide size-frequency distribution of particles extending from sub-wavelength-scale to sizes much larger than the wavelength, following a power-law size-frequency distribution (SFD) \cite{shoemaker1970}. In terms of the scattering properties, the grains that are much smaller than the wavelength, and are typically numerous and densely packed, form one scattering regime, the wavelength-scale particles form another regime, and the least numerous particles much larger than the wavelength form a third regime. In this work, we assume that the first regime of the tightly-packed small grains form a planar interface between free space and the surface of an atmosphereless planetary body, and wavelength-scale particles lay on the top of the plane surface (i.e., semi-infinite substrate). The third regime of particles much larger than the wavelength are not considered in this work, as the scattering properties tend to be dominated by the wavelength-scale particles due to the power-law SFD.

Extensive literature exists for particles of various shapes, sizes, and materials, both individual and as clouds in free space or embedded in a host material. The focus of this work is on non-symmetric wavelength-scale particles on a planar surface. Light scattering by spheres on or near surfaces has been solved analytically \cite{bobbert1986,videen1991,videen1992err}, while other simple shapes can be handled with semi-analytical (T-matrix) methods (e.g., \cite{doicu2008, mackowski2022}). However, realistic (irregular) particle shapes typically require the use of numerical (discretization-based) methods. While almost any methods can account for a substrate through direct discretization of its part (see, e.g., \cite{sun2009, albella2011}), the discrete dipole approximation (DDA) is one of a few methods that may include the substrate analytically through the modification of the Green's tensor describing the interaction between two dipoles \cite{taubenblatt1993, rahmani2001}. Unfortunately, such modification breaks the translation symmetry along the surface normal, which makes the corresponding codes \cite{loke2011, chaumet2021} significantly slower (and more so with increasing scatterer size). This performance gap along with a larger number of free parameters describing the scattering geometry explains why the simulations of light scattering by irregular particles on surfaces have been performed on a much smaller scale than for particles in free space, limited to studies of a few test cases instead of systematic variation of parameters.

The performance issue was solved in the ADDA code \cite{yurkin2015}, making the DDA calculation speed comparable to the free-space case and, in many cases, much faster than any alternative approach. The goal of this paper is to apply this simulation capability to rough polyhedral particles. Although the focus of this work is on microwave observations of regolith surfaces, the scale invariance (dependence of the scattering only on the size-wavelength ratio \citep{mishchenko2006}) makes the results informative at visible wavelengths as well. Thus, they are relevant not only in space science but in a variety of applications from understanding the light scattering by compact mineral dust particles on solid surfaces to Earth observations using remote sensing at any wavelengths that are comparable to the size of the scatterers. 

In Section \ref{sec:methods}, we provide the basic scattering theory underlying the definition of the investigated scattering properties, describe and justify the physical properties of the particles selected for this study, and explain how the computations were conducted. In Section \ref{sec:results}, we present the results of the computations beginning from example cases to explain the observation geometries and their effect on the scattering profiles in terms of general trends, then introduce the observed effects of the particle shape, size, and material on the scattering properties. In Section \ref{sec:discussion}, we discuss critically the practical implications of the results, and provide the final conclusions in Section \ref{sec:conclusions}. 

\section{Methods}
\label{sec:methods}
 
\subsection{Scattering theory}
\label{sec:theory}
Let us, first, stress that there is no universally accepted extension of the scattering quantities defined for the free-space scattering to the case of a particle near the substrate. In the vacuum (or a homogeneous host medium), a limited amount of scattering quantities (once computed) can be used to simulate detector signal in any scattering geometry for either a single particle or a suspension of them (at least, at small concentrations). In the case of a substrate, such desirable abstraction is not impossible, but is not yet fully developed and will require a significantly larger number of scattering quantities to cover all measurement scenarios \citep{yurkin2015,moskalensky2019}. While we discuss the extensions of scattering quantities to the case of a substrate and related ambiguities below, the final choice of quantities to present is partly a matter of convenience. Their total number is a compromise between generality and the ability to study them systematically in a single paper.

The simulated scattering properties include the extinction, scattering, and absorption cross sections and efficiencies, which can be used for calculating the single-scattering albedo, and the scattering matrix. The scattering matrix describes mathematically how the properties of the incident radiation change in the scattering process. If any particular electromagnetic radiation is given as a Stokes vector $\mathbf{I}$ with four elements $I$, $Q$, $U$, and $V$, where $I$ describes irradiance, $Q$ and $U$ describe the linear polarization properties, and $V$ describes the circular polarization (see, e.g., \cite{bohrman}), the scattered radiation is
\begin{equation}
	\mathbf{I}_\mathrm{sca} = \frac{\mathbf{F}}{k^2R^2} \mathbf{I}_\mathrm{inc}, 
 \label{eq:mueller}
\end{equation} 
where the subscript sca refers to "scattered", inc for "incident", $k$ is the vacuum wavenumber $2\pi \lambda^{-1}$ ($\lambda$ is the wavelength), $R$ is the distance from the scatterer to the observer, and $\mathbf{F}$ is the $4 \times 4$ Mueller scattering matrix (so that an element on row $p$, column $q$ is written as $F_{pq}$). This matrix has a well-known relation to the $2 \times 2$ amplitude scattering matrix elements ($S_1,...,S_4$) \cite{bohrman}. These expressions require additional care when the particle is placed near a plane substrate \citep{yurkin2015,yurkin2020}. When both incident and scattered directions are in the upper hemisphere (vacuum), they hold as is. When considering scattering into a non-absorbing substrate, the wavenumber $k$ in Eq. \ref{eq:mueller} needs to be replaced by that in the substrate. Moreover, some of the scattering directions may then be non-accessible (have zero scattering amplitude), corresponding to the phenomenon of total internal reflection. For scattering into an absorbing substrate, absolute value of the complex wavenumber in that medium is relevant and additional exponential decay with $R$ appears \citep{yurkin2015}. In the latter case, $\mathbf{F}$ can still be computed, but its physical relevance largely depends on the specific application.

The scattering matrix is defined with respect to the scattering plane, i.e., the one containing the incident and the scattered direction \cite{bohrman}. In the plots below, unless stated otherwise, we assume the scattering plane to contain the substrate normal (further denoted as vertical scattering plane). Thus, the azimuthal scattering angle in the laboratory reference frame coincides with that for the incidence (or differs by $180^\circ$). The two basic linear polarizations are then parallel and perpendicular to the scattering plane, which are commonly denoted as vertical (v) and horizontal (h) with respect to the substrate plane, respectively (e.g., \cite{ulaby2014, raney2021}). This definition is ambiguous for incidence normal to the substrate, but the difference between the two polarizations becomes irrelevant if azimuthal average is employed (as is the case in this paper). For 2D angular distributions, where the azimuthal scattering angle varies, the scattering plane and basic polarizations can be somewhat confusing. Regardless, the $F_{11}$ element (the only one that we consider in corresponding plots) has a clear physical meaning of scattered intensity for unpolarized incidence.


Other computed scattering properties include the extinction, absorption, and scattering efficiencies and cross sections (respectively $q_\mathrm{ext}$, $q_\mathrm{abs}$, and $q_\mathrm{sca}$; $C_\mathrm{ext}$, $C_\mathrm{abs}$, and $C_\mathrm{sca}$). The efficiencies describe how well the scatterer removes energy from the incident radiation through scattering, absorption, or both, whereas the cross sections refer to the respective parameters multiplied by the scatterer's geometric cross section ($C_\mathrm{G}$). Both extinction and absorption can be expressed through the integrals over the particle volume, which allow a natural generalization to the case of a substrate \citep{yurkin2015}. Extinction then accounts for the attenuation of both refracted and reflected plane waves (present without a particle). Thus, comparison of extinction efficiency or cross section between the free-space and substrate cases is potentially confusing due to various possible choices of a reference wave (which attenuation is of interest).

Even larger ambiguity is related to the scattering quantities. In the following, we employ the definition based on the optical theorem:
\begin{equation}
	C_\mathrm{sca} = C_\mathrm{ext} - C_\mathrm{abs}, 
\end{equation}
corresponding to the near-field scattered power (energy flow through the particle boundary). It can be contrasted to the far-field scattered power that can be expressed through the integral of some elements of $\mathbf{F}$ (depending on incident polarization) over all scattering directions \citep{moskalensky2019}. These two approaches are equivalent only for the free-space scattering or the case of a non-absorbing substrate; otherwise part of the scattered energy is absorbed in the substrate. The equivalence can be extended to the substrate with very weak absorption (if a fixed far-field distance is used, ensuring negligible attenuation). However, for all such substrates, scattered power integrated over only the upper hemisphere can be a more practically relevant quantity, which we do not discuss further. The cross sections and efficiencies are computed for two linear polarizations, v and h, defined above.


The computations were conducted using a wavelength of $2\pi$ (corresponding to $k=1$). The particles are then characterized by the size parameter $x=kr$, where $r$ is the particle radius, which fully determines the Mueller matrix $\mathbf{F}$ and the efficiencies. By contrast, the cross sections need to be divided by $k^2$ for applications requiring other wavelengths.

When a size-frequency distribution (SFD) is used, the scattering matrix is weighted based on the number of particles so that
\begin{equation}
	\displaystyle \left\langle \mathbf{F} \right\rangle = \frac{\int_{x_\mathrm{min}}^{x_\mathrm{max}} \mathbf{F}(x) N(x) \mathrm{d}x}{\int_{x_\mathrm{min}}^{x_\mathrm{max}} N(x) \mathrm{d}x},
\end{equation}
where $N(x)$ is the SFD of the size parameters. Because the size-parameter steps are not constant but increase in steps of 0.5 from to $x=0.5$ to 3 and in steps of 1.0 above 3, we use the nonuniform Simpson's rule \citep{shklov60} for the numerical evaluation of the integrals. The SFD choices are discussed further in Section \ref{sec:physprop}.

The incident Stokes vector depends on the simulated observation method: sunlight that illuminates planetary objects is effectively unpolarized (an incident-flux-normalized Stokes vector $\mathbf{I} = (1,0,0,0)$). In remote-sensing applications such as lidar and radar, the transmitted signal is typically either linearly or circularly polarized. Ground-based radar systems typically use circular polarization; e.g., Arecibo S-band radar transmitted right-handed circular polarization ($\mathbf{I} = (1,0,0,1)$ in the forward-scattering alignment), whereas the Miniature-Radio Frequency instrument on the Lunar Reconnaissance Orbiter used to transmit (and still receives) either linear polarization ($\mathbf{I} = (1,\pm 1,0,0)$). 

Let us assume first a circularly polarized incident signal: the opposite-circular (OC) and the same-circular (SC) polarized backscattering cross sections are, respectively, 
\begin{align}
\label{eq:NRCS}
\sigma_\mathrm{OC} & = \frac{2\pi} {k^2} [F_{11}(180^\circ) - F_{44}(180^\circ)] \quad \mathrm{and} \nonumber \\\sigma_\mathrm{SC} & = \frac{2\pi} {k^2} [F_{11}(180^\circ) + 2F_{14}(180^\circ) + F_{44}(180^\circ)], 
\end{align}
where $F_{14}(180^\circ)$ is typically much smaller than the other two elements for ensemble-averaged scattering matrices \citep{bohrman}. The ratio $\mu_\mathrm{C} = \sigma_\mathrm{SC}/\sigma_\mathrm{OC}$ is commonly known as the circular polarization ratio (CPR or SC/OC ratio). CPR is widely used in planetary, Earth, and lunar radar observations as a gauge to surface roughness; however, the connection between the observable parameters and the physical characteristics of the observed target or region is not trivial but many factors play a role in the observed values (e.g., \cite{ulaby2014,thompson2011,virkki2016}). Therefore, scattering simulations and laboratory experiments are crucial to help inform the interpretation of the observations. 

Here, the backscattering cross section is further normalized by the total geometric cross sections of the particles to obtain the normalized radar cross section (NRCS). For example, for the OC polarization (with $(180^\circ)$ omitted for brevity), the SFD-weighted NRCS is
\begin{align}
	\hat{\sigma}_\mathrm{OC} = \frac{\int_{x_\mathrm{min}}^{x_\mathrm{max}} 2\pi k^{-2} (F_{11}(r) - F_{44}(r))N(r) \mathrm{d}r}{\int_{x_\mathrm{min}}^{x_\mathrm{max}} N(r)C_\mathrm{G}(r) \mathrm{d}r} \nonumber \\
    = \frac{\int_{x_\mathrm{min}}^{x_\mathrm{max}} 2 (F_{11}(x) - F_{44}(x))N(x) \mathrm{d}x}{\int_{x_\mathrm{min}}^{x_\mathrm{max}} N(x)x^2 \mathrm{d}x},
\label{eq:sfd_weighted_nrcs}
\end{align}
as $x = kr$ and $C_\mathrm{G} = \pi r^2$. The latter definition is used for all particles (including polyhedrons), assuming $r$ to be the radius of a sphere with the same volume. For non-spherical particles, $C_\mathrm{G}$ is an effective quantity independent of particle orientation (not exactly equal to geometric cross section), but we use it consistently throughout the paper to calculate the efficiencies from the cross sections.

Linear polarization is often used in lidar and space-borne radar applications ($\mathbf{I} = (1,\pm 1,0,0)$). The same-linear (SL; in literature typically either hh or vv) polarized backscattering cross section is $\sigma_\mathrm{SL} = 2\pi k^{-2} [F_{11}(180^\circ) + F_{22}(180^\circ) \pm F_{12}(180^\circ)]$, where the choice of the sign depends on the scattering plane, and the orthogonal-linear (OL; either hv or vh) polarized backscattering cross section is $\sigma_\mathrm{OL} = 2\pi k^{-2} [F_{11}(180^\circ) - F_{22}(180^\circ)]$. The linear polarization ratio $\mu_\mathrm{L} = \sigma_\mathrm{OL}/\sigma_\mathrm{SL}$.

For ensembles that are statistically symmetric with respect to the incidence direction, e.g., orientation-averaged particles in free space, $F_{44} = F_{11} - 2F_{22}$ at backscattering  \citep{Mishchenko2002}, and thus:
\begin{equation}
 \mu_\mathrm{C} = \frac{2\mu_\mathrm{L}}{1-\mu_\mathrm{L}}
 \label{eq:cpr_lpr}
\end{equation}
 The applicability of this relation to particles on surfaces will be discussed in Section \ref{sec:backscat}.

\begin{figure}[ht]
	\centering
	\includegraphics[width=0.46\textwidth]{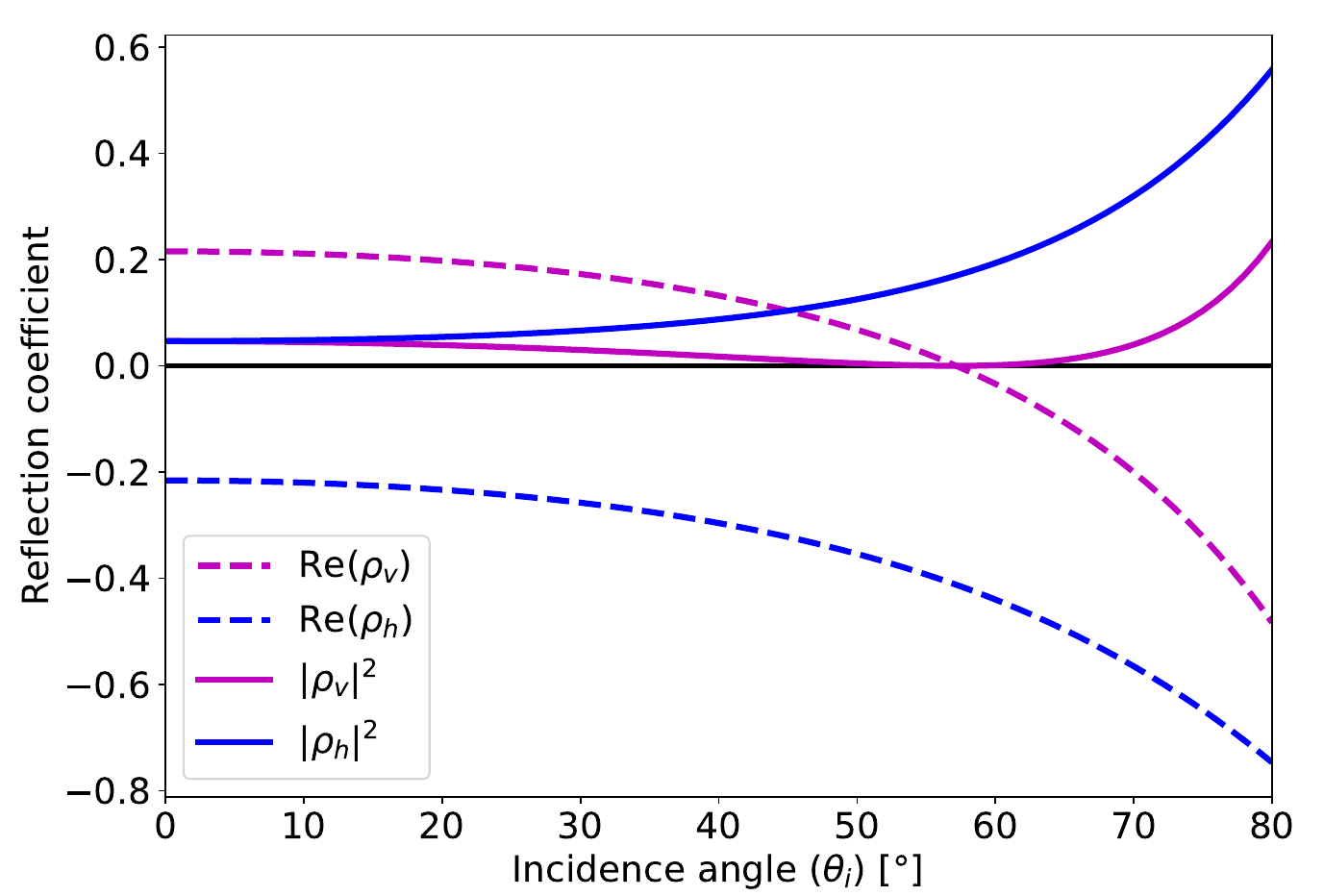}
	\caption{The reflection coefficients $\textrm{Re}(\rho_\mathrm{v})$ (magenta dashed line), $\textrm{Re}(\rho_\mathrm{h})$ (blue dashed line), $|\rho_\mathrm{v}|^2$ (magenta solid line), and $|\rho_\mathrm{h}|^2$ (blue solid line) as functions of the incidence angle. The Brewster angle is at $57.2^\circ$. }
	\label{fig:fresnel}
\end{figure}

The surface is assumed to be perfectly smooth, thus, simple Fresnel formulae apply to it without particles. Figure \ref{fig:fresnel} depicts the corresponding reflection coefficients for the used substrate refractive index ($m$) of $1.55 + 0.004$i.

To finalize this section and facilitate further discussion, let us summarize the general effects of a substrate on the scattering by particles. First, as elaborated above, it changes the scattering geometry (definition of scattering angles) and reduces the system symmetry (even for random particle ensembles) due to the presence of the preferential direction (substrate normal). Thus, one should not expect that the symmetry relations, e.g., for the elements of the Mueller matrix will be the same as that in the free space.

Second, the substrate globally modifies both the incident and scattered fields. The particle is subjected to the sum of direct and reflected plane waves, and the scattered field goes to infinity (above the substrate) either directly or after reflection from the substrate. Overall, for fixed incidence and scattering directions, there are four pathways: one directly scattered, two single-reflected, and one double-reflected, all of which interfere at the far-field region. Thus, we will use the term \textit{far-field interference} for these effects. Importantly, two incoming waves generally see different sides of the particle. The relative strengths of these pathways depend on the incidence angle through the Fresnel coefficients, but it is uncommon that all three substrate-related ones can be neglected.

Third, at each of the above pathways the particle scatters light differently than in free space, due to the interaction of the particle with itself through the substrate. The latter, which can be named \textit{near-field interaction}, is the most complicated part of the scattering problem (both conceptually and numerically). It can be described by scattered waves falling back on the particle after the reflection from the substrate. This interaction is expected to be more prominent for oblate particles lying on the surface than for general compact particles, due to larger average angles of reflection, and to increase with substrate refractive index. However, it is hard to make any prior estimate, and we are not aware of any qualitative indicators (features) in measurable quantities that are specific to this interaction. Thus, isolating and quantifying the near-field particle-substrate interaction is one of the goals of this paper.

\subsection{Rough polyhedral particles}

The polyhedral particle shape models were generated along the following steps:

\begin{enumerate}
\item A Delaunay triangulation for a unit sphere was generated using 4096 vertices that creates 8188 triangular face elements.
\item $N$ random vertices were selected as seed vertices, $N$ being the number of macro-scale faces the polyhedron should have.
\item The closest seed vertex was searched for all vertices, and the angle $\delta$ was calculated between the vertex vectors. 
\item The length of each vertex vector was scaled using $(2\cos(\delta))^{-1}$ to build a tangent plane through each of the seed vertices. 
\item Surface roughness was added by multiplying all the new vertex coordinates by a Gaussian-random number with a mean of 1.0 and a standard deviation of 0.02.
\item The axis ratio (longest to shortest axis) was allowed to vary up to 1.8, or else the shape was regenerated.
\item 16 realizations of irregular dodecahedrons (12 faces) and irregular icosahedrons (20 faces) were generated; however, not all realizations were used for every case. Each realization was then discretized into voxels.
\end{enumerate}

\begin{figure}[ht]
	\centering
\includegraphics[width=0.46\textwidth]{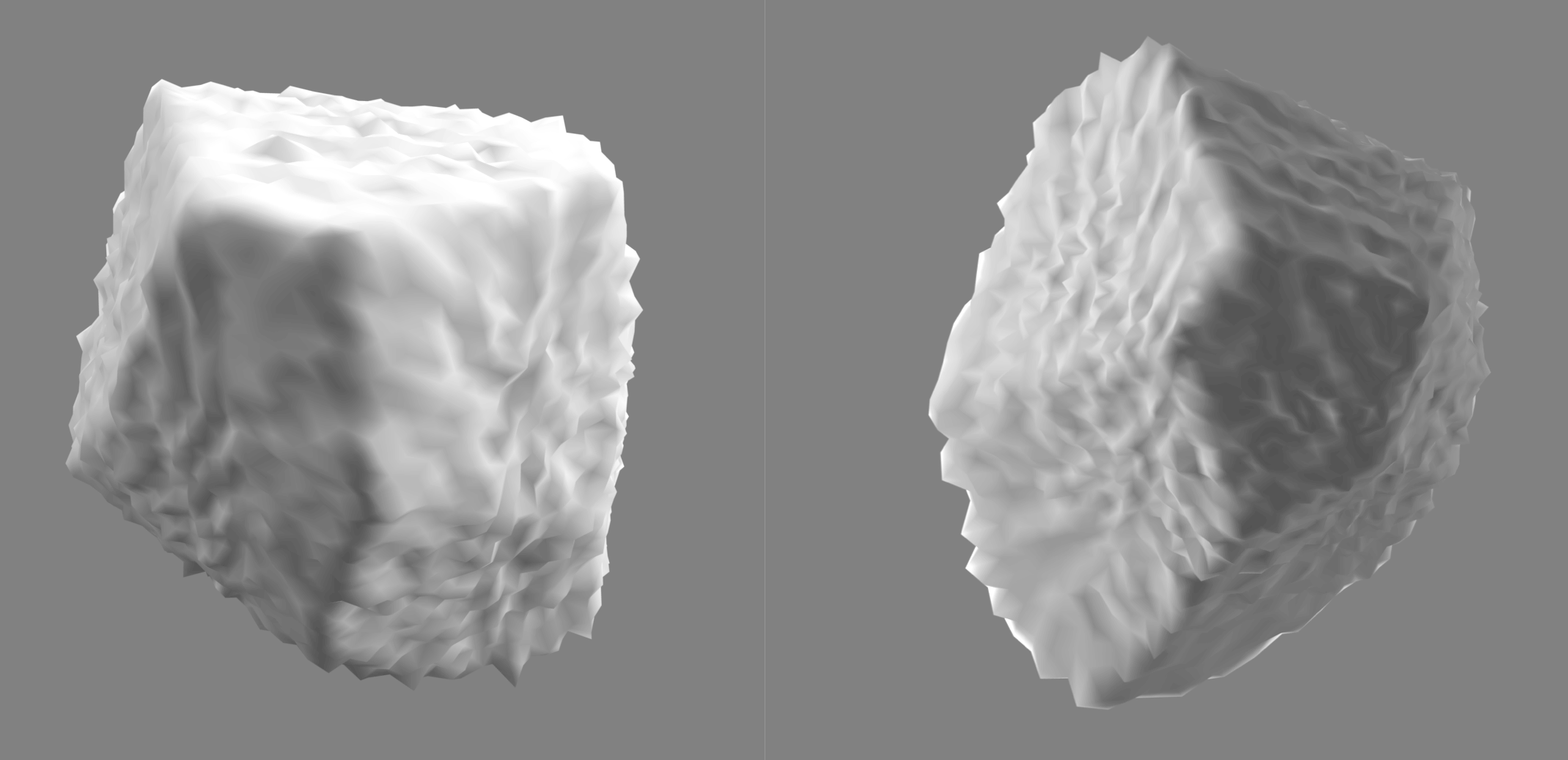}
	\caption{Two example particles: a 12-face polyhedral particle (on the left) and a 20-face polyhedral particle (on the right).}
	\label{fig:exampleparticle}
\end{figure}

The number of faces determines statistically the axis-ratio distribution of the polyhedral particles, because a larger number of faces generates statistically rounder particles, whereas particles with a small number of faces tend to be more angular and are overall more stochastic. Figure \ref{fig:exampleparticle} shows examples of two polyhedral particles, with either 12 or 20 faces, with surface roughness added but before the discretization. The difference in roundness is not evident for these individual realizations in single orientations. 

The area of individual faces is larger for a smaller number of faces, which could affect the coherence of the scattering phases at large sizes. However, for small sizes, the specific shape of the particle plays a minor role. A simple test for a polyhedron with $x=8$ illuminated from a direction normal to a face (at a $60^\circ$ inclination to the normal of the surface), compared to another illumination angle and with different levels of surface roughness, did not show a clear backscattering enhancement (see Appendix \ref{sec:roughness}). Therefore, no significant contribution is expected in the size range selected for this study. However, the incident EM radiation propagation through the faces could cause forward-diffraction in the internal fields.

The role of the surface roughness is an intriguing light scattering question on its own, but it likely plays a small role in the selected range of size parameters and non-symmetric particle shapes. A couple of tests to investigate the role of the surface roughness are presented in Appendix \ref{sec:roughness}. Thus, the treatment of the surface roughness is purposefully simplistic. It is uncorrelated between consecutive vertices and depends on the number of vertices, because the distance between two consecutive vertices affects the slopes of the triangular elements within the face. The roughness is not uniform through the face, but the center of the face close to the seed vertex has larger density of vertices and smaller absolute height variations. For the vertices farther from the center, the absolute height variations increase and the vertices are sparser. Also, the multiplication using the vertex vector instead of the seed vertex unit vector causes the roughness spikes to point increasingly away from the seed vertices. The effect is not noticeable when the height variations are small and especially when the particles are discretized into voxels, but could be an issue with greater standard deviations or for very large particle sizes (e.g., in future studies).  

\subsection{Physical properties of the particles}
\label{sec:physprop}

Beyond their morphology, the physical properties of the individual particles include electric properties (permittivity) and the size range. For particular materials, the SFD and the packing density of the material also play roles. Furthermore, depending on the wavelength and the size range of the particles, the effective permittivity of the material could be different from that of the grains, e.g., for tightly packed particles much smaller than the wavelength. 

Here, two different relative permittivities are investigated to demonstrate their role in the scattering properties, specifically, $\epsilon_r=4.7 + 0.016$i ($m=2.17 + 0.004$i) and $\epsilon_r=7.8 + 0.09$i ($m=2.79 + 0.0155$i), which represent a range of values for most solid, rocky metal-poor geologic materials \cite{campbell1969}. For the powdered regolith substrate, we use $\epsilon=2.4 + 0.012$i ($m=1.55 + 0.004$i) based on the permittivity measurements of \cite{campbell1969} for various rock and meteorite powders at a range of densities, and assuming a bulk density of 1260~kg~m$^{-3}$ estimated for (101955) Bennu in \cite{chesley2014}. Because the computations require the refractive index as a parameter, we will refer to the materials hereafter by the refractive index. We note that previous studies of light scattering by polyhedrons (recently, e.g., by \cite{muinonen2023}), typically consider lower refractive indices due to the main focus on optical observations.

Regarding size scales, the particles are assumed to be roughly comparable to the wavelength, specifically in the range $x \in [0.5,8]$. The radii and size parameters of the polyhedrons are determined using spheres of the same volume (the same as for $C_\mathrm{G}$ above). For instance, the sphere-equivalent particle diameters would be $4.01x$~cm at S-band wavelengths of 12.6~cm, or $1.13x$~cm at 3.55-cm X-band wavelengths. For optical applications, at an average visual wavelength of 0.55~$\mu $m, the diameters would be $0.18x~\mu$m. 

\begin{figure}[ht]
  \centering
    \includegraphics[width=0.4\textwidth]{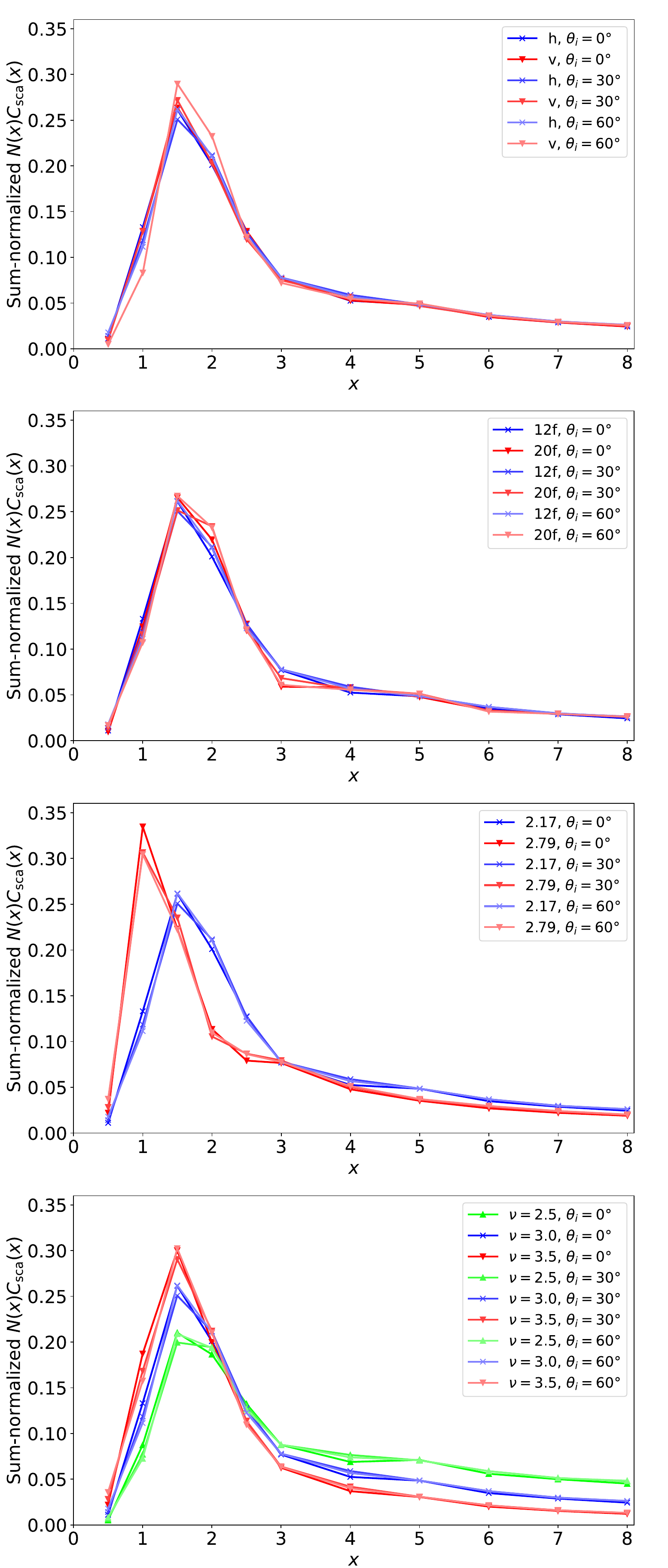}
    \caption{The sum-normalized SFD-weighted scattering cross sections for the polyhedrons as a function of the size parameter for different incidence angles (shades), when physical properties are varied: on the top, two orthogonal polarizations (horizontal in blue and vertical in red); second, two numbers of faces (12 faces in blue and 20 faces in red); third, two refractive indices ($m=2.17 + 0.004$i in blue and $m=2.79 + 0.0155$i in red), and on the bottom, the SFD power-law index $\nu$ is varied from 2.5 (green) to 3.5 (red), when $N(x) \propto x^{-\nu}$. The blue lines are the same in all panels with horizontal polarization, 12-face polyhedrons, $m=2.17 + 0.004$i, and $\nu=3.0$.}
    \label{fig:wCsca}
\end{figure}

All investigated size-frequency distributions are based on the power law: $N(x) \propto x^{-\nu}$, where $\nu \in [2.5, 3.5]$. As shown in \cite{virkki2016}, when $\mathrm{Re}(m) > 2$ and the particle SFD has a power-law distribution $N(x) \propto x^{-3}$, the scattering properties are dominated by particles with size parameters 1--3, based on the shape of $N(x)C_\mathrm{sca}(x)$. The same result was confirmed to apply here for both orthogonal polarizations (Fig. \ref{fig:wCsca}; here $N(x)C_\mathrm{sca}(x)$ is normalized by its sum over the range of included size parameters). We find only minor discrepancies between the horizontal (h) and the vertical (v) polarization as a result of the surface reflection even at large incidence angles. When $\nu \geq 3$, the scattering contribution at $x=8$ and above is less than 5 \% of the total scattering cross section, and therefore has little impact in the SFD-weighted results. 
 

\subsection{Computational methods}
\label{sec:computations}
The scattering properties of the rough polyhedral particles were computed in the resonance regime ($x \in [0.5,8]$, with individual cases up to $x=9$) using the code ADDA \cite{yurkin2011}. For convenience, ADDA was also used for spheres, although much more efficient alternatives -- analytical solutions -- exist for this case  \cite{bobbert1986,videen1991}. The computations for the larger particles were conducted using the supercomputers at the CSC --~IT Center for Science, whereas the computations for the smaller particles could be handled using a regular laptop. 
The same computational approach has been previously used for rough flat cylinders on a surface \cite{riskila2021}.

For the ADDA computations, the particles were discretized into dipoles, or numerically cubical voxels, with sizes much smaller than the wavelength. The discretization was done using the Point-Inside-Polyhedron (PIP) code included in the ADDA package \citep{schuh2007}. Typically ADDA computations are considered to be accurate when the number of dipoles per wavelength (dpl) is at least $10|m|$. For non-spherical particles with intentionally pseudorandom shapes, the criterion may potentially be looser. The particle discretization was done with different voxel resolutions ensuring that the dpl $> 9|m|$ in all the presented simulations. For the small particles ($x \leq 3$), which contribute the most in the power-law-weighted cases, the number of dipoles along any coordinate axis was at least 37 to ensure that their shape is properly resolved. The spherical reference particles were computed using 32 dipoles per wavelength ($14.7|m|$ for $m=2.17+0.004$i). We used the Filtered coupled dipoles (FCD) formulation of the DDA and the Modified quasi-minimal residual (QMR2) iterative solver with threshold of $10^{-4}$ for the relative residual norm. The accuracy of ADDA is discussed further in Appendix \ref{sec:accuracy}.

Some of the most interesting questions regarding the surface particles are how their scattering properties differ from those in free space, how the scattering properties of irregular particles differ from those of spherical particles, and what are the practical implications for better interpretation of observations. 

\begin{figure}[ht]
  \centering
\includegraphics[width=0.46\textwidth]{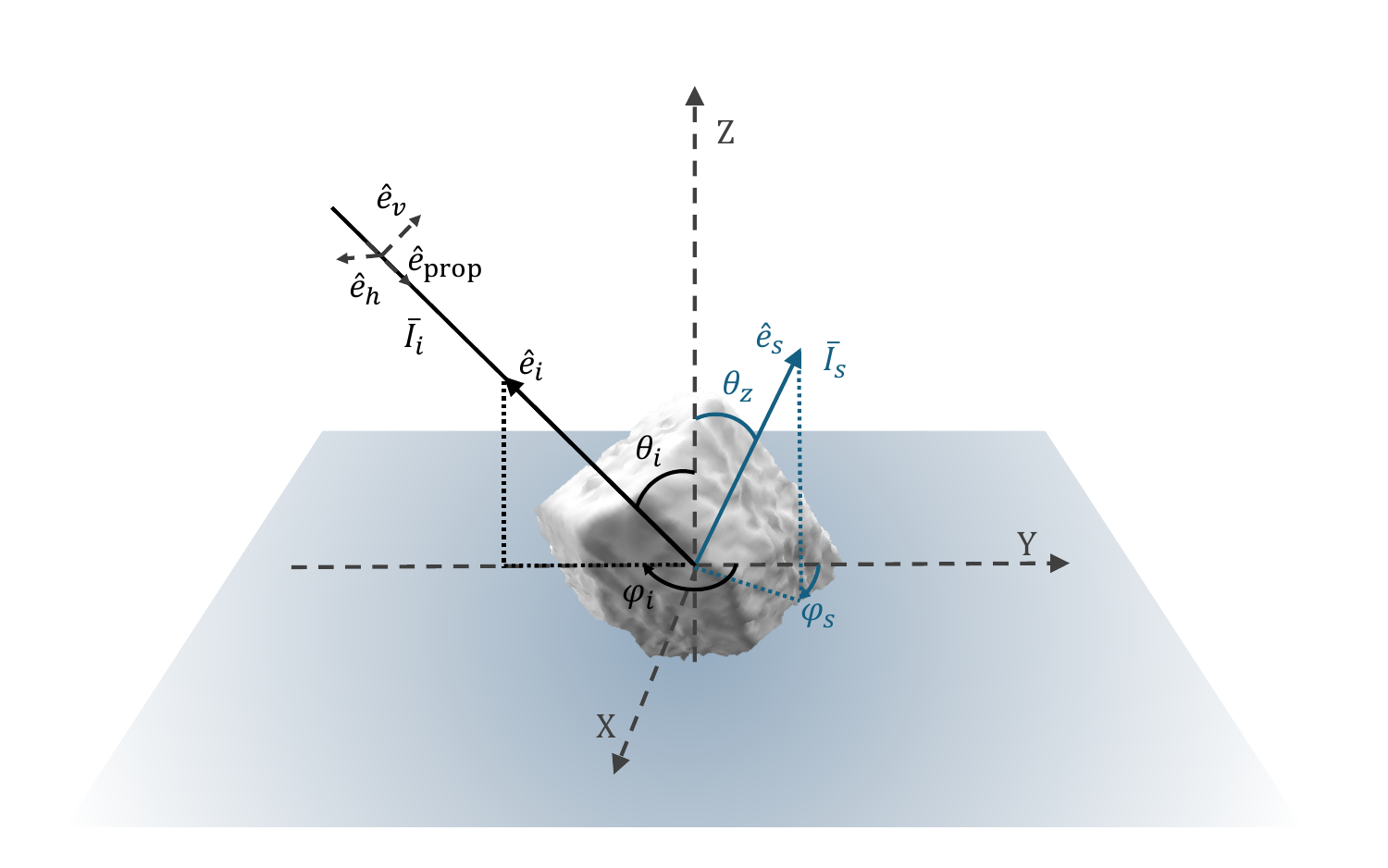}
\caption{The schematic of the scattering geometry illustrating the relevant angles for incident (black) and scattered (blue) vectors. The former is defined using a zenith angle $\theta_i$ and an azimuth angle $\varphi_i$, and the latter using $\theta_z$ and $\varphi_s$. In the vertical scattering plane, $|\varphi_i-\varphi_s|=180^\circ$.}
\label{fig:geometry}
\end{figure}

\begin{figure}[ht]
  \centering
\includegraphics[width=0.46\textwidth]{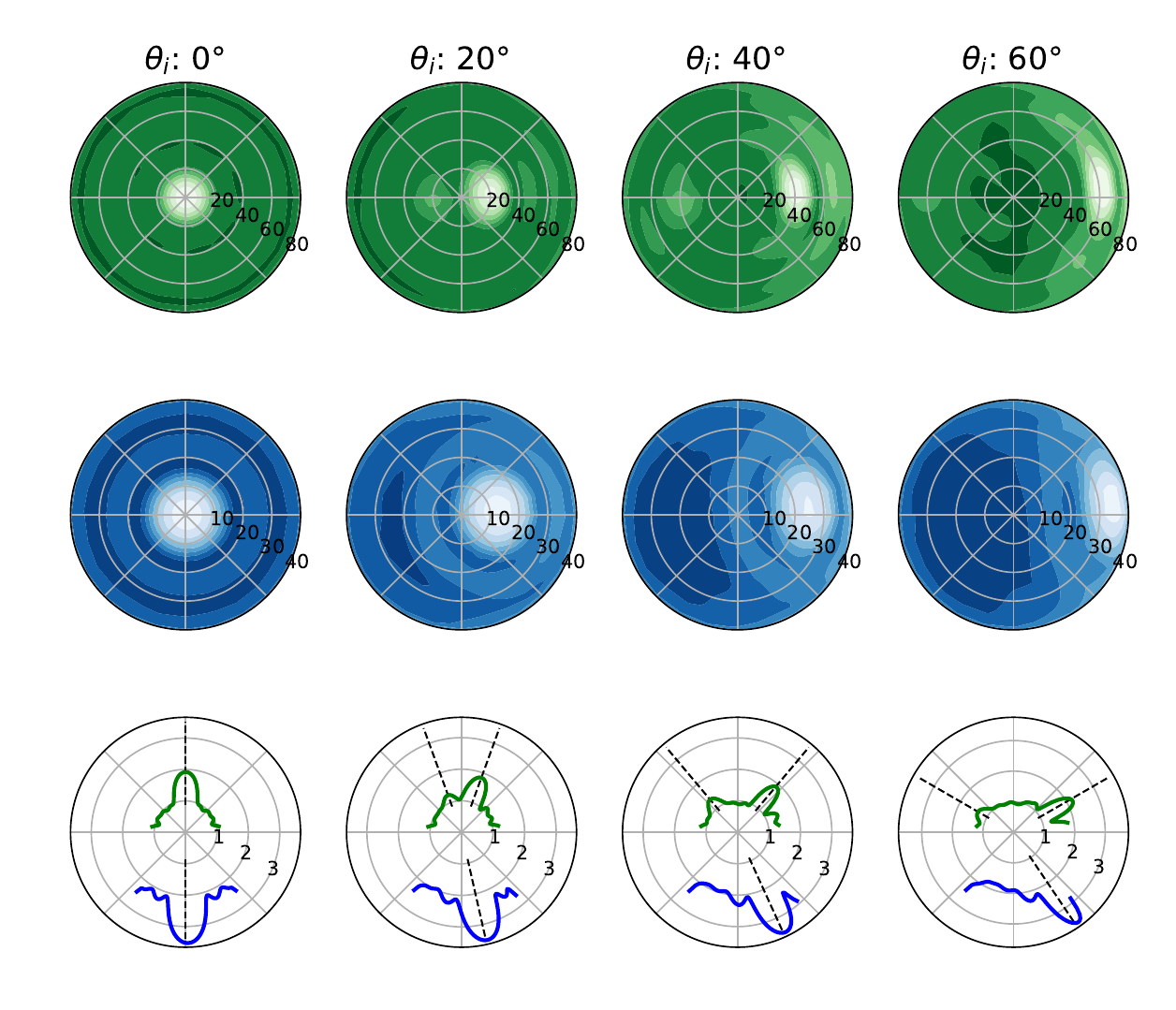}
	\caption{The angular distribution of $\log_{10}F_{11}$ in the zenith and nadir views for azimuthally and ensemble-averaged scattering matrices of 20-face polyhedrons, when $x=6$ and the incidence angle increases from 0$^\circ$ to 60$^\circ$ (columns from left to right). The top row shows the zenith view and the middle row shows the nadir view, so that the vertical scattering plane corresponds to the horizontal middle line and the incidence approaches from the left (above the page) towards right (into the page) for all $\theta_i > 0^\circ$. The values by each concentric circle display the angular extent: $[0^\circ,80^\circ]$ from zenith in the zenith view and $[0^\circ,40^\circ]$ from nadir in the nadir view. The lighter shades depict larger values. The full range of values can be inferred from the bottom row, which shows $\log_{10}F_{11}$ as a function of zenith angle (with $\theta_z=0^\circ$ in top center) for both the zenith (upper hemisphere) and the nadir (lower hemisphere) views in the vertical scattering plane.}
	\label{fig:F20_x6_az}
\end{figure}

When EM radiation is scattered by a wavelength-scale (resonance regime) particle in free space, the corresponding geometry can be simply defined based on the directions of the incident and emitted rays, the scattering properties can be integrated cylindrically about the forward-scattering direction, and the 1,1-element of the scattering matrix typically displays a pronounced forward-diffraction peak and a number of oscillations as a function of the scattering angle. For a particle on a surface, it is sensible to use a laboratory reference frame, where the zenith angle aligned with the normal of the surface is the reference direction (Fig. \ref{fig:geometry}). While the zenith is considered to be in the direction of the surface normal above the surface, the nadir is the opposite direction below the surface. The incidence angles are, thus, defined as zenith angles $\theta_i$ of the incident radiation. In this scheme, we investigate $\theta_i$ in steps of 10$^\circ$ from 0$^\circ$ to 60$^\circ$ while the incident azimuth angle $\varphi_i$ increments by 45$^\circ$ in eight steps over a full rotation. For computational reasons, the normal-incidence case is computed at $\theta_i = 0.8^\circ$ to be able to vary $\varphi_i$ the same way as for other $\theta_i$. It would also be possible to obtain the results for all values of $\varphi_i$ at normal incidence by special postprocessing of the full angular distribution for a single simulation; however, the selected approach was considered more convenient in this work and has a negligible difference to an ideal normal-incidence case. Incidence angles greater than 60$^\circ$ are not included, because then it is less realistic to neglect particle-to-particle interactions (at least, shadowing) even at small surface coverage; however, the grazing incidence would be an interesting case for a future study. For the scattered azimuth angle $\varphi_s$, we use steps of 15$^\circ$ for the polyhedrons and 5$^\circ$ for spheres. For a particle on a surface, two diffraction peaks can emerge, one at the reflection angle and another at the refraction angle with respect to the plane surface (see Fig. \ref{fig:F20_x6_az} and its description in the next section), which has to be considered in the angular averaging. 

All scattering profiles for polyhedrons presented in Section \ref{sec:results} are averaged over the eight incident azimuthal orientations so that the rotation of the scattering plane is considered, which is effectively equal to averaging over eight rotation phases of the particle around the Z-axis for a fixed incident propagation vector. Each orientation is simulated independently. The orientations can be considered comparable to additional realizations; however, considering uncertainties, there could be a difference in the covariance of the scattering matrix elements averaged over eight rotation phases of a single realizations in contrast to that of eight realizations in a single orientation. Investigating this possible difference in detail is not in the scope of this paper. With regards to the results, the ensemble and azimuthal averaging together ensure good satisfaction of scattering-matrix symmetries (discussed in Section \ref{sec:obs_geom}).

The computations for polyhedrons in free space were done in the particle reference frame using $81\times32$ orientations, so that for each of the 81 different incidence directions (relative to the particle) 32 different scattering planes were considered (equivalent to the rotation of the particle along this direction). These 81 orientations and corresponding weights for averaging were determined using the "optimal cubature on the sphere" scheme, recommended in \citet{penttila2011}.  The reference direction for the scattering angles is, then, the forward-scattering direction. The ensemble averages for particles in free space include twelve particle realizations.

For optical and synthetic aperture radar (SAR) applications, both phase (in radar applications also bistatic) angle, $\alpha$, and incidence angle can be meaningful, whereas in Earth-based planetary radar observations, $\theta_i$ is a primary parameter. Therefore, the scattering as a function of both phase (or bistatic) and incidence (or zenith) angle are investigated. The computations have been conducted in the laboratory reference frame so that $\theta_i$ is easy to extract. The phase angle has to be calculated using the difference of incidence and scattering (emission) angles so that $\alpha = \arccos(\hat{\mathbf{e}}_i \cdot  \hat{\mathbf{e}}_e)$, where $\hat{\mathbf{e}}_i$ is the unit direction vector of the illumination source viewed from the scatterer, and $\hat{\mathbf{e}}_e$ is the unit direction vector of the emission (or the observer). The scattering matrices for the same phase angles are not necessarily equal at different zenith angles even in the vertical scattering plane, while outside this plane additional rotational terms would be required. For simple visualization purposes, we limit ourselves to averaging over two scattering directions in the vertical scattering plane, corresponding to the same phase angle. This is unambiguous for employed $\theta_i$ and $\alpha \leq 25^\circ$. Note also that emission refers here only to the imminently scattered energy from the incidence. Thermal or other passive emission is not included here.

Another key parameter is the particle height above the surface. Since particles are not allowed to intersect the plane surface for computational reasons, the distance from the plane surface was adjusted individually for each particle so that the lowest voxel is as close to the surface as is computationally possible. The effect of the distance to the surface is out of the main scope of this paper, but is illustrated for an example case in Appendix \ref{sec:distance}. 

The particles could lay on the plane surface in physically unrealistic positions (e.g., standing off-balance); however, the position was considered irrelevant regarding the average scattering properties as long as the particle is in the immediate vicinity of the surface.

\section{Results}\label{sec:results}

First, the general features of the scattering profiles due to the observation geometries are shown using selected example cases. Second, the effects of particle shape and size are systematically illustrated. Third, the effect of material is shown. Apart from that subsection, the default refractive index is $2.17 + 0.004$i. Fourth, the average scattering properties for different SFDs are presented, approaching a realistic comparison to ground-truth observations. Finally, the significance of including the substrate in the computations, in contrast to the free-space case, is discussed in Section \ref{sec:surfacevsfree} with focus on the backscattering. All presented results for polyhedral particles are ensemble- and azimuthally averaged as described unless we explicitly mention otherwise. For spherical particles neither averaging is necessary.

\subsection{Observation geometries}
\label{sec:obs_geom}

\begin{figure*}[htb]
  \centering
	\includegraphics[width=0.9\textwidth]{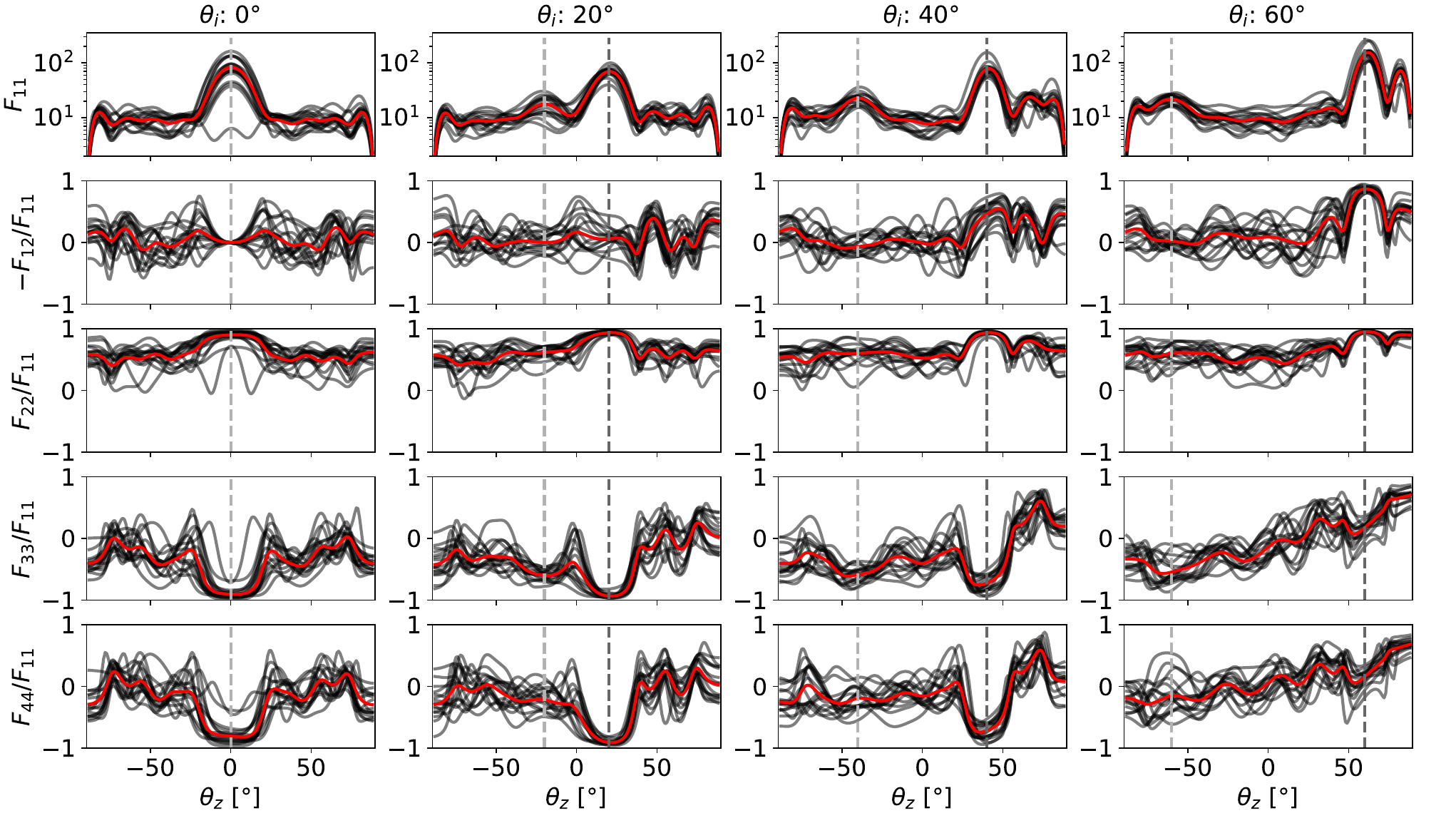}
	\caption{The azimuthally averaged scattering matrix elements $F_{11}$, $-F_{12}/F_{11}$, $F_{22}/F_{11}$, $F_{33}/F_{11}$, and $F_{44}/F_{11}$ in the vertical scattering plane as a function of zenith angle for 20-face polyhedrons, when $x=6$ and the incidence angle increases from 0$^\circ$ to 60$^\circ$ (columns from left to right) both for individual particle realizations (solid dark gray) and the ensemble average (solid red). The vertical dashed lines show the incidence (and thus, backscattering) angle (light gray) and the reflection angle (dark gray).}
	\label{fig:Poly_x6_pol}
\end{figure*}

Figures \ref{fig:F20_x6_az} and \ref{fig:Poly_x6_pol} display two example cases of scattering profiles for 20-face polyhedrons when $x=6$, $m=2.17+0.004$i, and the incidence angle increases from 0$^\circ$ to 60$^\circ$ in steps of 20$^\circ$. Both examples show azimuthally averaged profiles. Figure \ref{fig:F20_x6_az} displays the full angular distribution of $\log_{10}F_{11}$ for an ensemble average over 16 particle realizations, while Fig. \ref{fig:Poly_x6_pol} displays the scattering profiles of individual realizations averaged over eight azimuthal orientations as well as the ensemble average in the vertical scattering plane. A specular peak with a forward-diffraction-like shape emerges both at the reflection and transmission angles (the latter is present only in the nadir hemisphere in Fig. \ref{fig:F20_x6_az}). The reflection peak becomes more prominent with increasing $\theta_i$, following the increase of the reflection coefficient (see Fig. \ref{fig:fresnel}). Both sets also show a modest backscattering peak in $F_{11}$.

In the subsurface hemisphere (i.e., nadir view), we omit the data for zenith angle from 90$^\circ$ to 140$^\circ$ (the angle of total internal reflection for the substrate refractive index), as was explained in Section \ref{sec:methods}. Although the computed values for these values are not zero (they are instead very large for slightly absorbing substrate), they correspond to rapidly decaying scattered waves that are hardly relevant for applications. Moreover, the focus of the paper (and of the discussed applications) is on the top hemisphere (zenith angle of 0$^\circ$ to 89$^\circ$). 

Note also that polarization effects are discussed only in figures such as Fig. \ref{fig:Poly_x6_pol}, where the use of a vertical scattering plane avoids any ambiguities related to the definition of incident polarization. The minor remaining asymmetry with respect to the vertical scattering plane (the horizontal middle line in the first two rows of Fig. \ref{fig:F20_x6_az}) is due to imperfect averaging, discussed in Appendix \ref{sec:accuracy}.

\begin{figure*}[htb]
  \centering
	\includegraphics[width=0.9\textwidth]{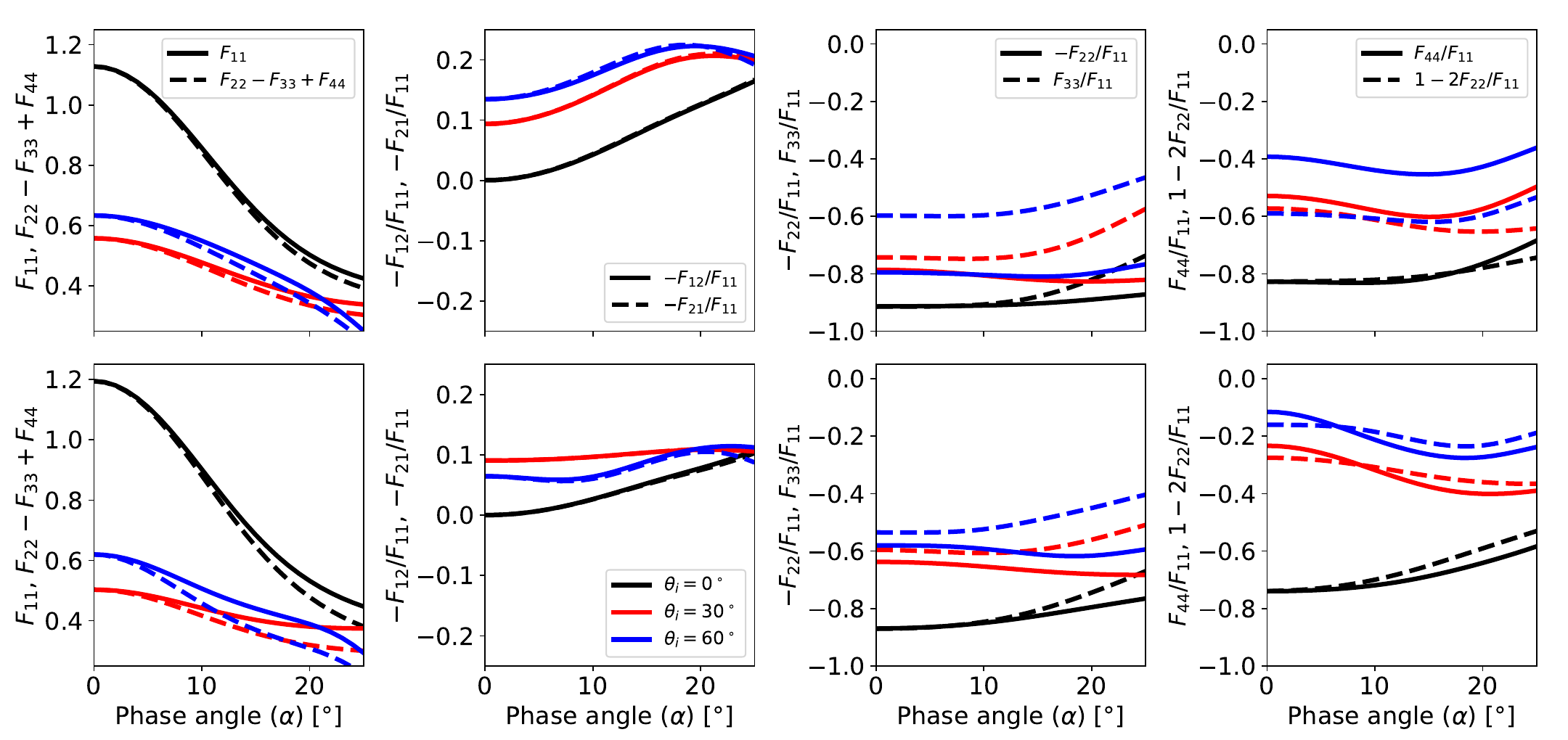}
	\caption{The azimuthally and ensemble-averaged scattering matrix elements $F_{11}$, $-F_{12}/F_{11}$, $-F_{22}/F_{11}$, $F_{33}/F_{11}$, and $F_{44}/F_{11}$ in the vertical scattering plane as a function of phase angle (angular extent from the backscattering direction) for 20-face polyhedrons (top row) and 12-face polyhedrons (bottom row) with a power-law SFD ($x^{-3}$ for $x \in [0.5,8.0]$) when $m=2.17 + 0.004$i. The dashed and solid lines illustrate symmetries known to apply for random ensembles in free space, but not necessarily near a substrate. The incidence angle varies from $0^\circ$ to $60^\circ$ (line colors as labeled in the legend of the bottom row's second panel from the left).}
	\label{fig:Poly12vs20_phase}
\end{figure*}

Figure \ref{fig:Poly12vs20_phase} displays the scattering matrix elements as a function of phase angle, which is commonly used for particles in free space. This representation provides a closer view to the values near the backscattering direction ($\alpha = 0^\circ$) and how they evolve as the angular extent increases. The graphs also reveal more subtle similarities in the backscattering. For instance, any scattering system (including that with substrate) must satisfy $F_{11}  + F_{33} = F_{22}  + F_{44}$  and $F_{12} = F_{21}$ at backscattering, which follows from $S_4 = -S_3$ for the amplitude scattering matrix \citep{vandehulst}. We have verified this fact for each simulated particle and, thus, for all averages. However, we additionally observe $F_{33} \approx -F_{22}$, which is under the above condition equivalent to $F_{44} \approx F_{11} - 2F_{22}$ (with the exception of 20-face polyhedrons at $\theta_i=60^\circ$). As discussed in Section \ref{sec:theory}, this is a known fact for averaging in free space and for normal incidence with the substrate (black lines in Fig. \ref{fig:Poly12vs20_phase}), while its satisfaction for other cases may indicate weak particle--substrate interaction. Another indication of the latter is that $F_{12} \approx F_{21}$ for non-zero $\alpha$, but we do not further discuss $F_{21}$, nor $F_{34}$ or $F_{43}$.

In terms of applications, there is notably little variation in the polarization elements within $20^\circ$ from the backscattering direction. Thus, polarimetric measurements of a surface as a function of phase angle are not likely to show much variation due to surface particles. 

\subsection{Effect of shape and size}
\label{sec:sh_size}
In this section, the effects of particle size and shape are shown in more detail. Key questions are how the scattering profiles of the polyhedral particles differ from those of spherical particles in terms of both the irradiance distribution and polarization properties? And what role does the size parameter play in the visible features? 

\begin{figure*}[htb]
  \centering
	\includegraphics[width=0.9\textwidth]{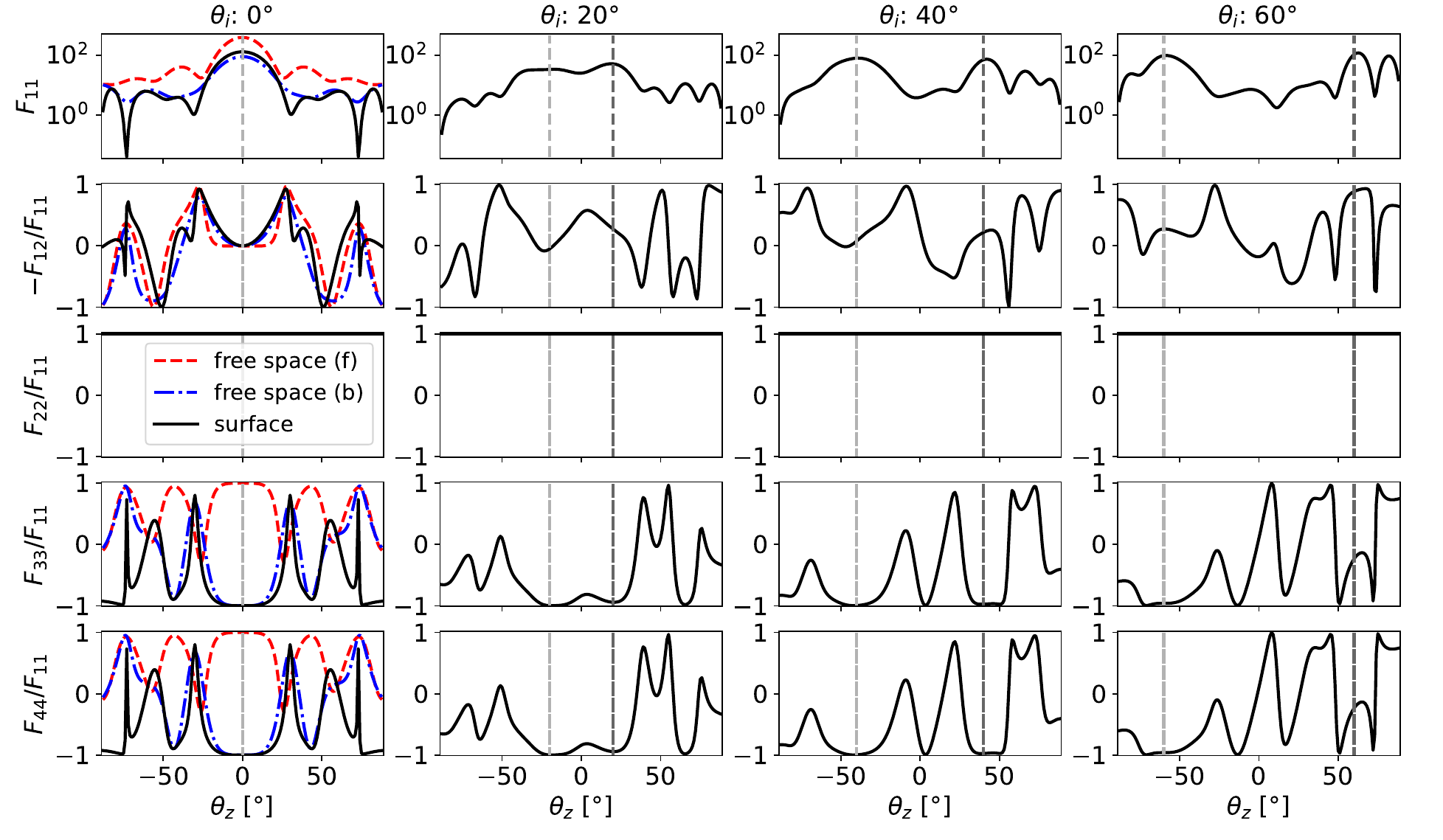}
	\caption{The same as Fig. \ref{fig:Poly_x6_pol} but for a spherical particle on a surface (black line) and, for reference, in free space for zenith angles with backscattering at 0$^\circ$ (blue dash-dotted line) and forward scattering at 0$^\circ$ (red dashed line), when $x=6$ and the incidence angle increases from 0$^\circ$ to 60$^\circ$ (columns from left to right). The vertical dashed lines show the incidence/backscattering angle (light gray) and the reflection angle (dark gray). The element $F_{22} = F_{11}$ and $F_{33} = F_{44}$.}
	\label{fig:Surfsph_pol}
\end{figure*}

Let us begin from scattering profiles for spherical particles. Figure \ref{fig:Surfsph_pol} displays the diagonal elements of the scattering matrix and $-F_{12}/F_{11}$ in the vertical scattering plane, when $x=6$. The scattering profile of a sphere in free space is included for reference to the panels where $\theta_i = 0^\circ$, so that both forward and backward profiles are displayed. Note however, that for this $\theta_i$ (and $x$), the zenith hemisphere for surface scattering is effectively the backscattering one for the free-space case. The forward scattering in free space is very different, apart from the overall shape and number of peaks, determined by $x$. Specifically,  $F_{44}/F_{11}=1$ for the latter, but $-1$ for the backscattering configuration ($\theta_i = \theta_z = 0^\circ$ in the surface case), which agrees with the general symmetry considerations for any geometry that is rotationally symmetric with respect to the propagation direction \citep{Mishchenko2002}. By contrast, the backscattering for the free space (all elements of the Mueller matrix) is surprisingly similar to the surface case for $\theta_i = 0^\circ$, especially for not too large $|\theta_z|$, which indicates that both far- and near-field effects of substrate are small (as defined in Section \ref{sec:theory}). The elements $F_{33}$ and $F_{44}$ are equal for all cases, which follows from the fact that the scattering system is symmetric with respect to the scattering plane. 

For $\theta_i \neq 0^\circ$, the neighborhood of the backscattering direction still resembles the free-space case. The specular-reflection direction is related to the free-space forward scattering, but only after the latter is adjusted for the reflection coefficient (two reflected pathways in Section \ref{sec:theory} ). More specifically, the elements $S_1$ and $S_2$ of the amplitude scattering matrix should be multiplied by $\rho_\mathrm{h}$ and $\rho_\mathrm{v}$, respectively. For small $\theta_i$ we have $\rho_\mathrm{v} \approx -\rho_\mathrm{h}$ (see Fig. \ref{fig:fresnel}) implying reduced $F_{11}$, roughly the same $-F_{12}/F_{11}$ and inverse sign of $F_{33}/F_{11}$. For $\theta_i = 60^\circ$, which is close to the Brewster angle, $\rho_\mathrm{v} \approx 0$, implying $F_{12} \approx -F_{11}$ and $F_{33} \approx 0$. This simplified description ignores the far-field interference of directly scattered and reflected pathways. It is especially obvious for $\theta_i = 20^\circ$ and explains the non-negligible value of $F_{33}/F_{11}$ for $\theta_z = \theta_i = 60^\circ$. However, we do not see any specific features of the near-field particle-substrate interaction. 



\begin{figure*}[h!]
  \centering
\includegraphics[width=0.88\textwidth]{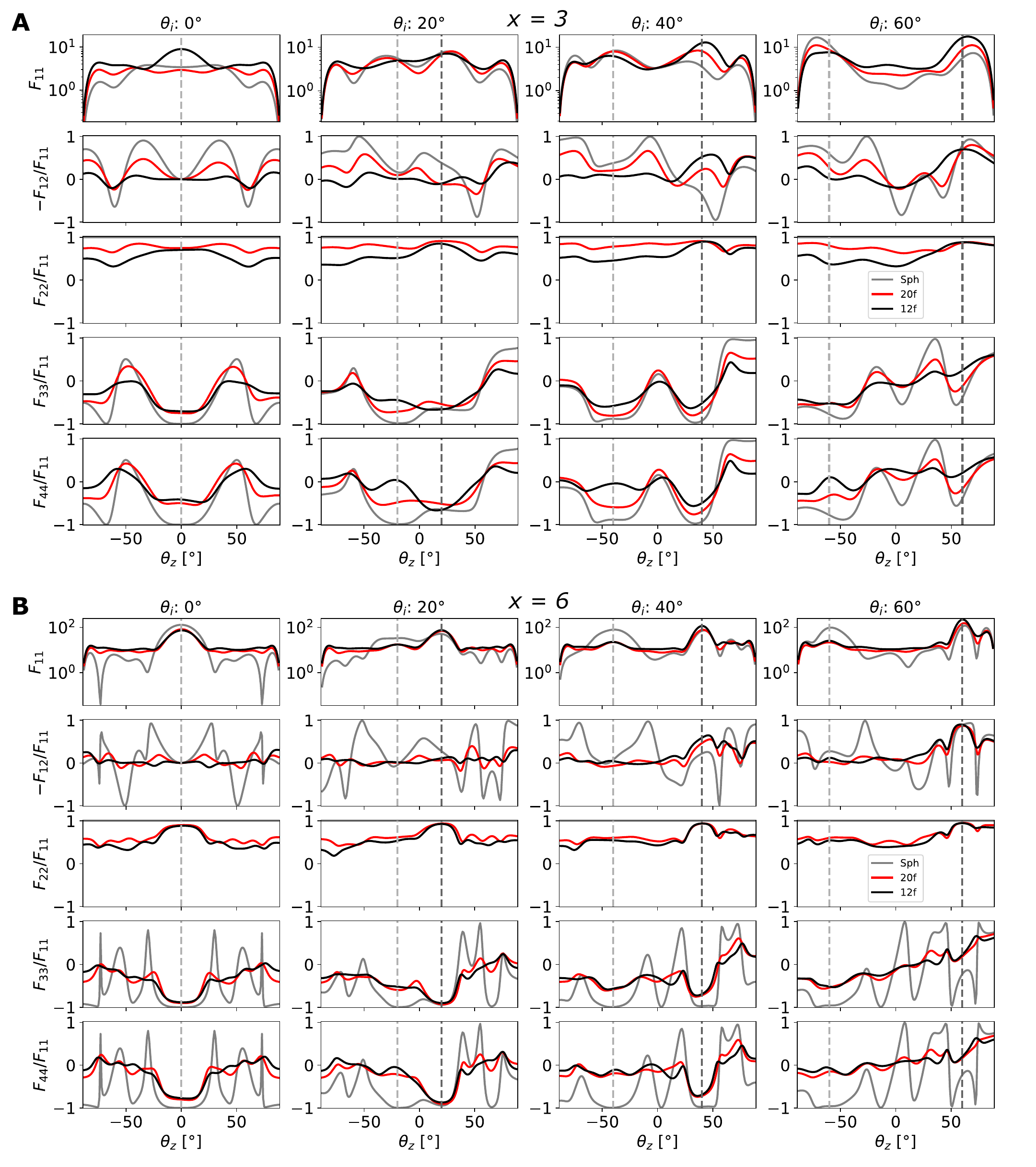}
	\caption{The azimuthally and ensemble-averaged scattering matrix elements $F_{11}$, $-F_{12}/F_{11}$, $F_{22}/F_{11}$, $F_{33}/F_{11}$, and $F_{44}/F_{11}$ in the vertical scattering plane as a function of zenith angle for spherical particles (solid gray line), 20-face polyhedral particles (red line) and 12-face polyhedral particles (black line), when $x=3$ (A; top five-row mosaic) or $x=6$ (B; lower five-row mosaic) and the incidence angle increases from 0$^\circ$ to 60$^\circ$ (columns from left to right). The vertical dashed lines show the incidence (light gray) and the reflection angles (dark gray).}
	\label{fig:Poly_x6_12vs20_pol}
\end{figure*}

Further, we compare the spherical particles to the polyhedral particles in Fig. \ref{fig:Poly_x6_12vs20_pol}; part of the data shown there (for $x=6$) is the same as in Figs. \ref{fig:Poly_x6_pol} and \ref{fig:Surfsph_pol}. First obvious difference is that the results for polyhedrons have smaller oscillations with $\theta_z$, as expected from all the involved averaging. Second, there is clear difference in the minimal values of $F_{44}/F_{11}$. For spheres, it is close to  $-1$ both at backscattering and, for $\theta_i \leq 40^\circ$, at specular reflection, while for polyhedrons it is larger for both of these regions, and significantly so at backscattering. Overall, for polyhedral particles, the minimal value of $F_{44}/F_{11}$ increases with the incidence angle as is also visible for individual realizations in Fig. \ref{fig:Poly_x6_pol}.

\begin{figure}[ht]
 \centering
	\includegraphics[width=0.45\textwidth]{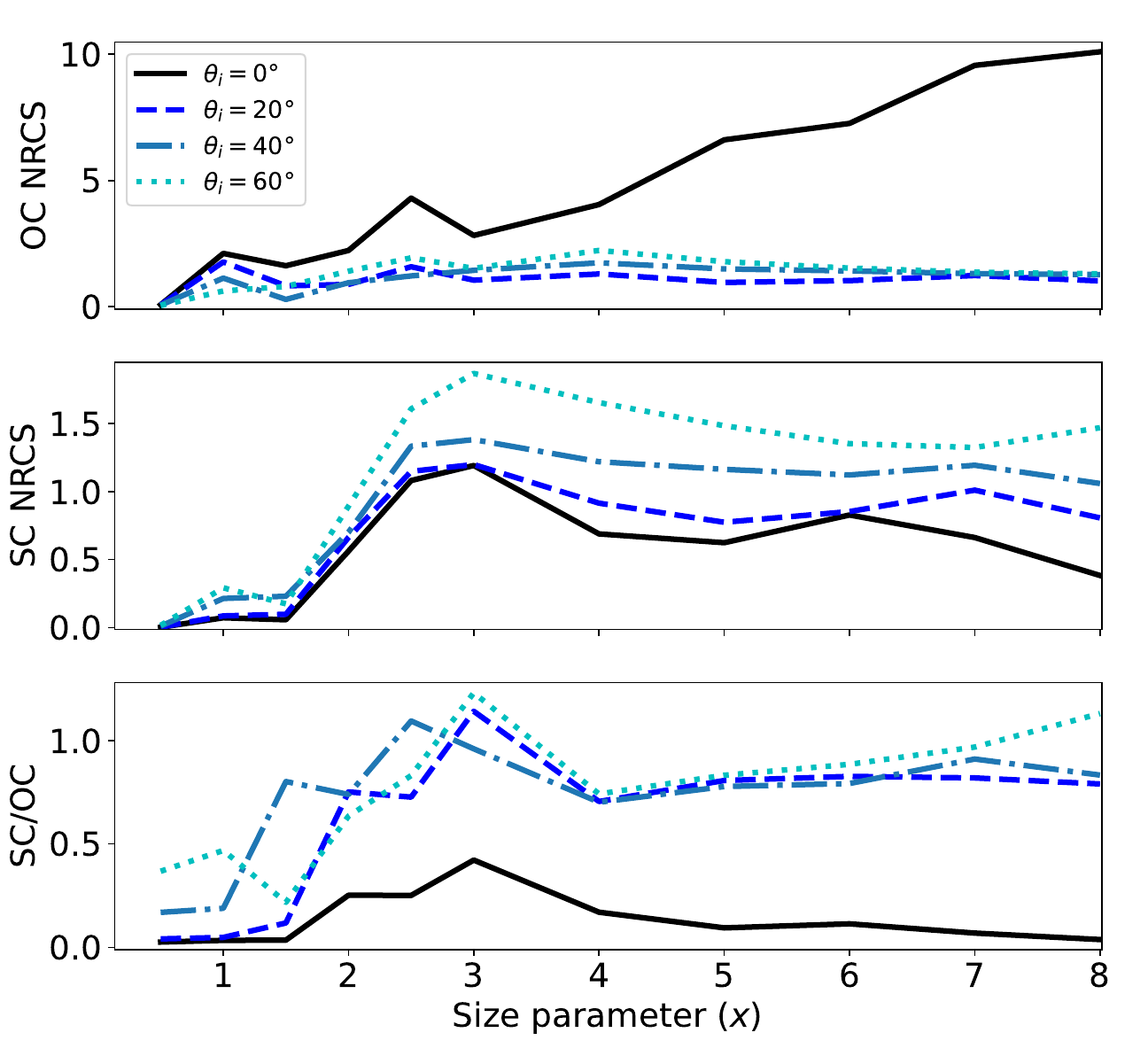}
	\caption{The backscattering OC NRCS, SC NRCS, and the SC/OC ratio of 12-face polyhedral particles ($m=2.17+0.004$i) on a surface as a function of size parameter using incidence angles from $0^\circ$ to $60^\circ$ (line styles as labeled).}
	\label{fig:SurfPoly12_CP_x}
\end{figure}

Third, $F_{22} \neq F_{11}$ and $F_{33} \neq F_{44}$ for polyhedrons, since individual particles are not mirror-symmetric with respect to the scattering plane. It is mostly prominent at backscattering directions for $\theta_i \neq 0^\circ$ and $x=6$, where $|F_{44}| \ll |F_{33}|$ and $F_{22}/F_{11} \approx 1/2$. For this specific case, we also observe $F_{33} \approx -F_{22}$.  For $x=3$, we observe similar behavior for 12-face but not for 20-face polyhedral particles; the latter case is roughly between the 12-face polyhedrons and spheres. The small 20-face polyhedrons have relatively small surface area of facets; thus, we can expect small effect of these facets on the scattering properties and the averaged result to be close to that of a sphere. Thus, it is understandable that both polyhedrons with $x=6$ are significantly different from a sphere even after averaging, but it is surprising that there is so little difference between them. Still, when a power-law SFD is used (giving larger weight to smaller particles) observable differences are expected. 

By contrast, the specular reflection peak has many similarities between all shapes. Specifically, we always observe $F_{22} \approx F_{11}$ and $F_{33} \approx F_{44}$, while for $\theta_i = 60^\circ$ we get the same Brewster-angle limits as discussed above ($F_{12} \approx -F_{11}$ and $F_{33} \approx 0$). Moreover, the condition $F_{33} \approx 0$ is even better satisfied for polyhedrons, probably due to smaller effect of directly scattered light. The latter is notably larger for $x=3$, which may explain the remaining differences between various shapes at specular reflection for this size.

Figure \ref{fig:SurfPoly12_CP_x} illustrates the dependence of the backscattering OC NRCS, SC NRCS, and the SC/OC ratio on the size parameter at different values of $\theta_i$ for the 12-face polyhedrons, derived from the azimuthally and ensemble-averaged scattering matrix elements using Eq. \ref{eq:NRCS}. The relatively larger dependence of OC NRCS on $x$ at normal incidence is related to the specular diffraction peak. It is intriguing that, at $\theta_i \neq 0$, the backscattering properties only weakly depend on $x$ when $x>3$. This fact also affects the SFD-weighted case discussed further in Section \ref{sec:backscat}. 

It is not clear whether the enhancement at normal incidence would be similarly present in a real-life experiment with a larger number of particles interfering in the far-field. In this case we expect more pronounced and narrower specular peak even for smaller particles (based on preliminary tests). Regardless, the backscattering by the substrate at normal incidence likely dominates the backscattering by the particles, which makes this issue less relevant in remote-sensing applications. 

\begin{figure*}[htb]
 \centering
	\includegraphics[width=0.9\textwidth]{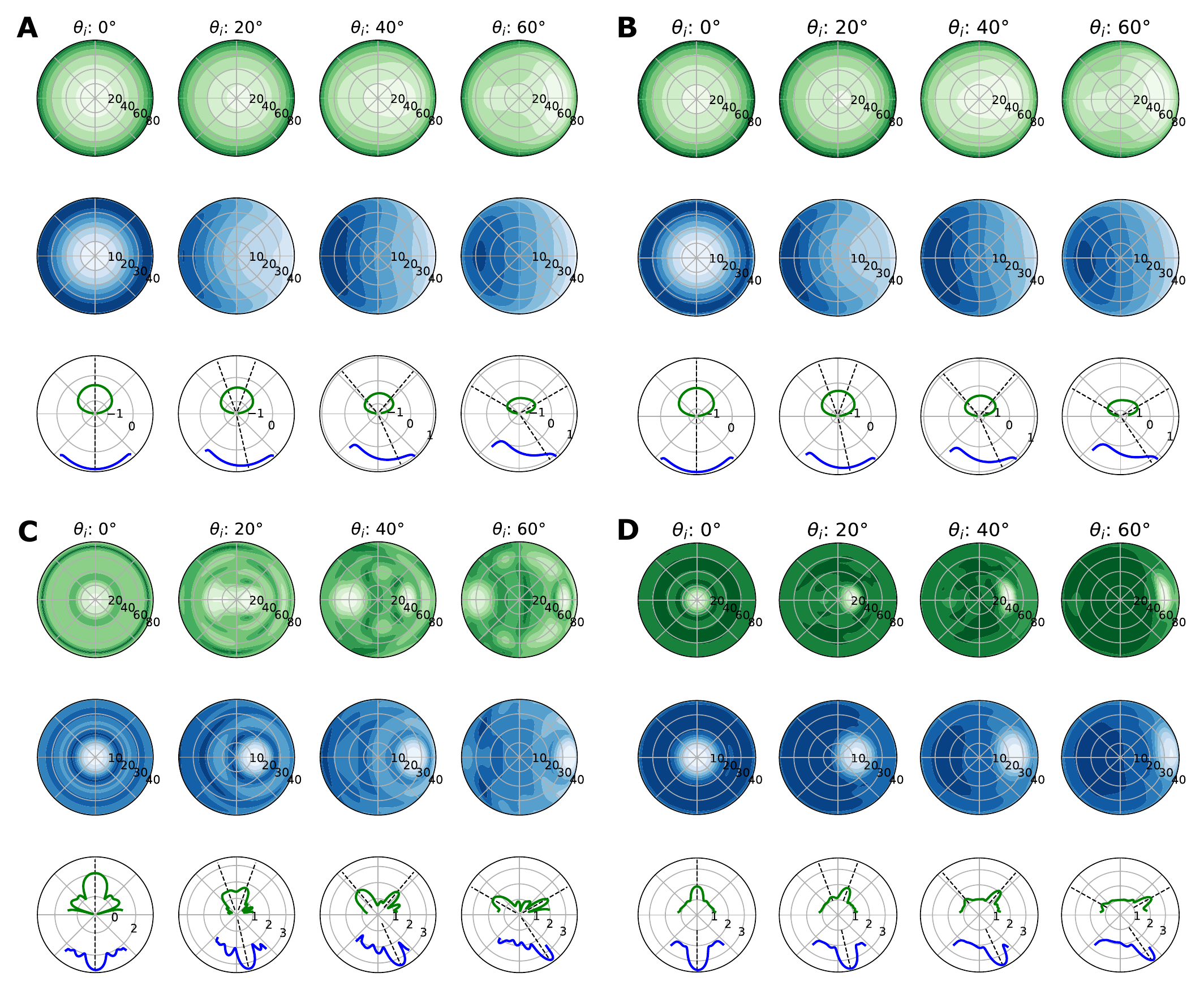}
	\caption{The angular distribution of $\log_{10}F_{11}$ in the zenith and nadir views as well as for the vertical scattering plane (as in Fig. \ref{fig:F20_x6_az}) for spherical particles (sets A and C) and azimuthally and ensemble-averaged 12-face polyhedral particles (B and D), when $x=1$ (A and B) and $x=6$ (C and D).}
	\label{fig:SurfSphPoly12}
\end{figure*}

Figure \ref{fig:SurfSphPoly12} shows the angular distribution of $\log_{10}F_{11}$ in the zenith and nadir views for scattering by a spherical particle compared to a polyhedron. At $x=1$, the shape plays a negligible role and the scattering is completely featureless, whereas at $x=6$ the sphere results in a rich angular structure, which is largely averaged out for polyhedrons. Comparison of Fig. \ref{fig:SurfSphPoly12}D to Fig. \ref{fig:F20_x6_az} confirms the similarity between 12- and 20-faces polyhedrons for this size. 

\subsection{Effect of material}

\begin{figure*}[htb]
  \centering
\includegraphics[width=0.9\textwidth]{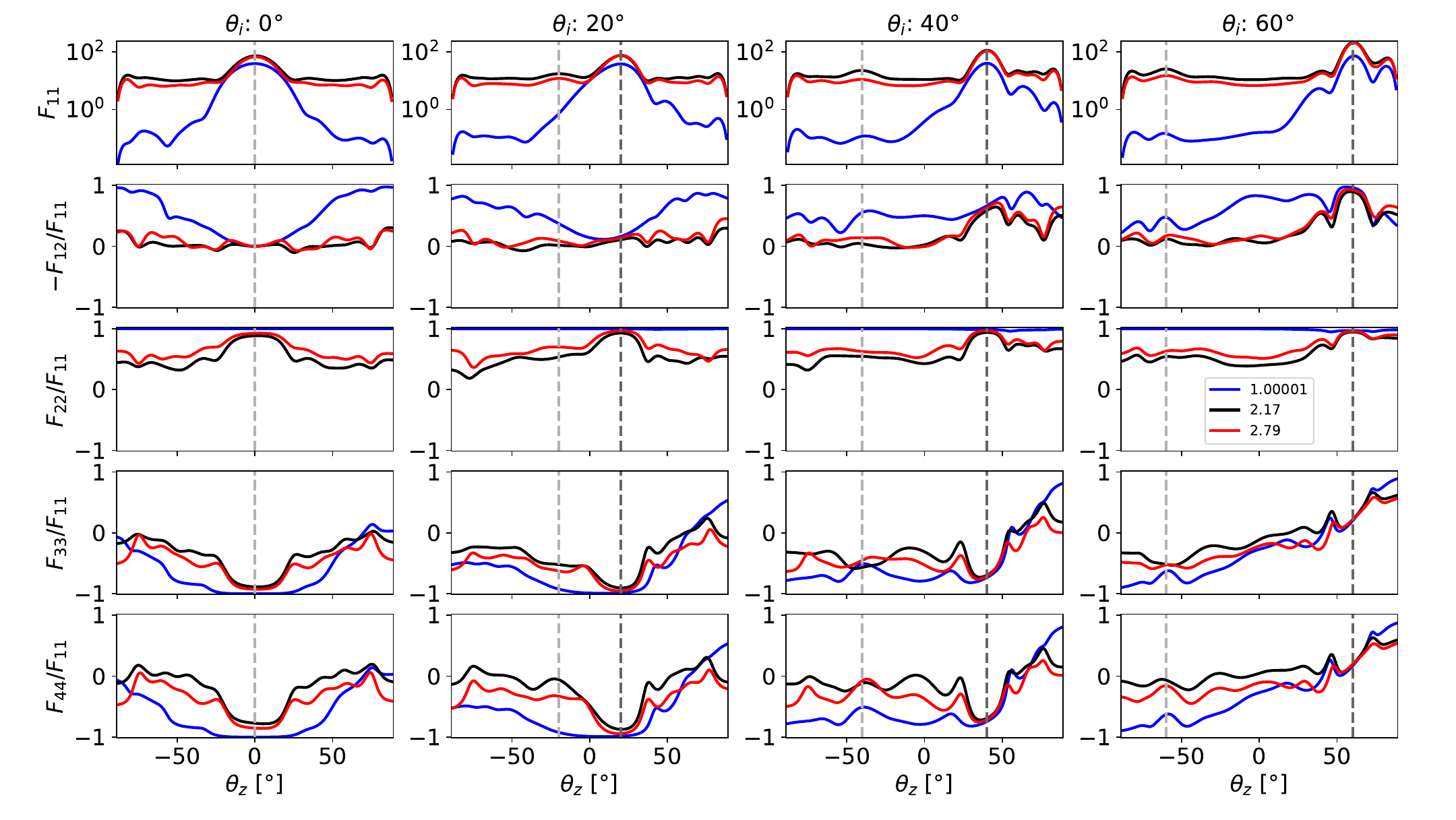}
	\caption{The scattering matrix elements $F_{11}$, $-F_{12}/F_{11}$, $F_{22}/F_{11}$, $F_{33}/F_{11}$, and $F_{44}/F_{11}$ in the vertical scattering plane as a function of zenith angle for 12-face polyhedrons with  and $m = 2.79 + 0.0155$i (red line), $m = 2.17 + 0.004$i (black line), and $m = 1.00001$ (blue line), when $x=6$ and the incidence angle increases from 0$^\circ$ to 60$^\circ$ (columns from left to right). The vertical dashed lines show the incidence (light gray) and the reflection angles (dark gray). The $F_{11}$ elements for $m = 1.00001$ have been multiplied by $10^8$ to facilitate comparison to the other two cases.}
	\label{fig:Poly_x6_refri}
\end{figure*}

Figure \ref{fig:Poly_x6_refri} shows the scattering matrix elements for the 12-face polyhedrons for two typical values of the refractive index and a reference case of $m = 1.00001$ (no absorption). The results for the high-refraction cases are based on ensemble averages of 12--16 realizations, whereas for the reference case only ten realizations were used due to the shape playing a minor role with such a low refractive index. Similar to the comparison of the number of polyhedron faces (Fig. \ref{fig:Poly_x6_12vs20_pol}B), there is relatively little effect due to the refractive index of the polyhedrons between $m = 2.79 + 0.0155$i and $m = 2.17 + 0.004$i. Notably, both of them have $F_{22}/F_{11} \approx 0.5$ and $F_{12}/F_{11} \approx 0$ except for the vicinity of the specular direction.

For comparison, the case of $m = 1.00001$ can be described by Rayleigh--Gans--Debye approximation (RGD), which effectively turns off both the near-field interaction of the particle with the substrate and the depolarization at the particle (hence, $F_{22} = F_{11}$). However, the scattering of the two incident polarizations are still not equivalent due to different reflection coefficients for the substrate. The forward-scattering (diffraction) peak has a comparable width (determined by $x$), but more rapid decay towards backscattering by a factor $\mathcal{O}(x^{-3})$.  Note, however, that measuring this forward peak is hardly feasible against the background of specular reflection from the substrate (that is not included in the plots).

Not surprisingly, such an artificial case significantly differs from two conventional ones, although there are also some similar features for all refractive indices. The largest similarity is at the forward-scattering directions, since all the cases satisfy the symmetry considerations discussed above. The $m = 1.00001$ case satisfies them best due to negligible effect of both directly scattered light and near-field particle-substrate interaction. By contrast, at backscattering directions for $\theta_i > 0^\circ$, the $m = 1.00001$ case has values of polarization elements roughly between that for spheres and polyhedrons with $\textrm{Re}(m)>2$.

\subsection{Size-averaged backscattering profiles} 
\label{sec:backscat}

\begin{figure*}[htb]
  \centering
	\includegraphics[width=0.9\textwidth]{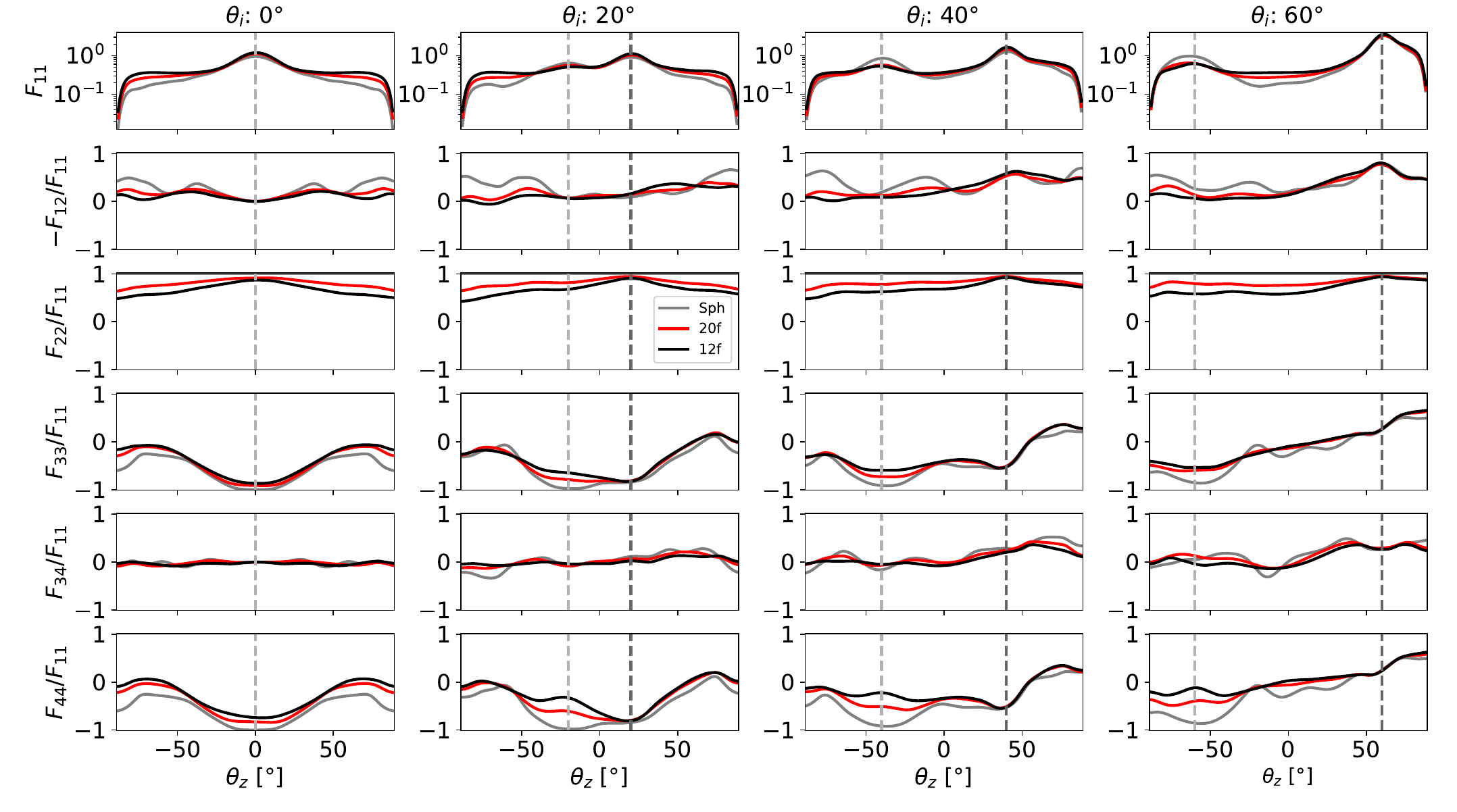}
	\caption{The block-diagonal azimuthally and ensemble-averaged scattering matrix elements for a power-law SFD $N(x) \propto x^{-3}$, where $x \in [0.5,8]$, for different shaped particles: spherical (gray solid line), 20-face polyhedrons (red line) and 12-face polyhedrons (black line). The incidence angle increases from 0$^\circ$ to 60$^\circ$ (columns from left to right).}
	\label{fig:Poly12vs20_SFD_pol}
\end{figure*}

\begin{figure*}[htb]
  \centering
	\includegraphics[width=0.9\textwidth]{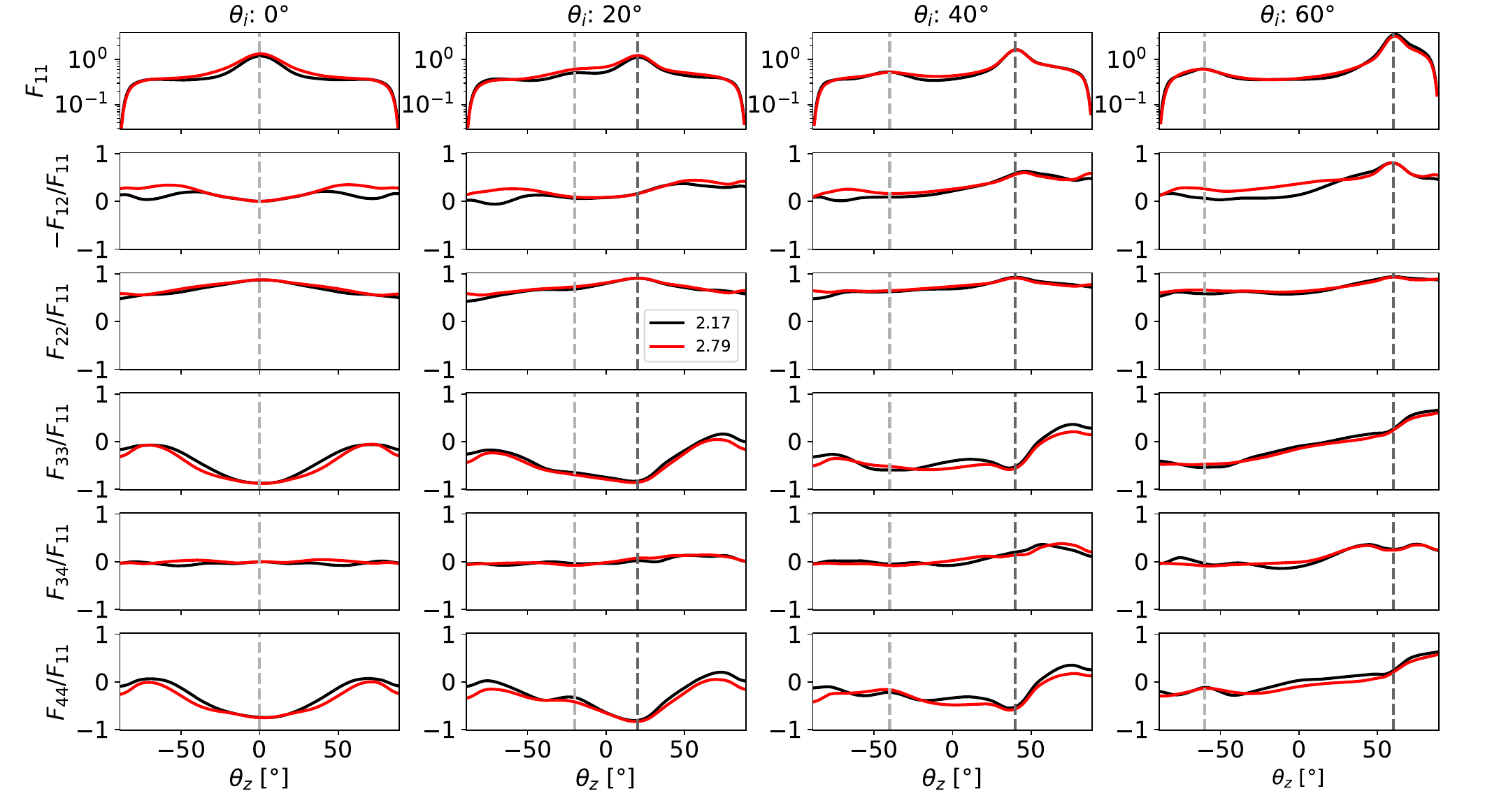}
	\caption{The same as Fig. \ref{fig:Poly12vs20_SFD_pol} but for 12-face polyhedrons with different refractive indices: $2.17+0.004$i (black line) and $2.79+0.0155$i (red line).}
	\label{fig:Poly217vs279_SFD_pol}
\end{figure*}

\begin{figure*}[ht]
	\centering
\includegraphics[width=0.9\textwidth]{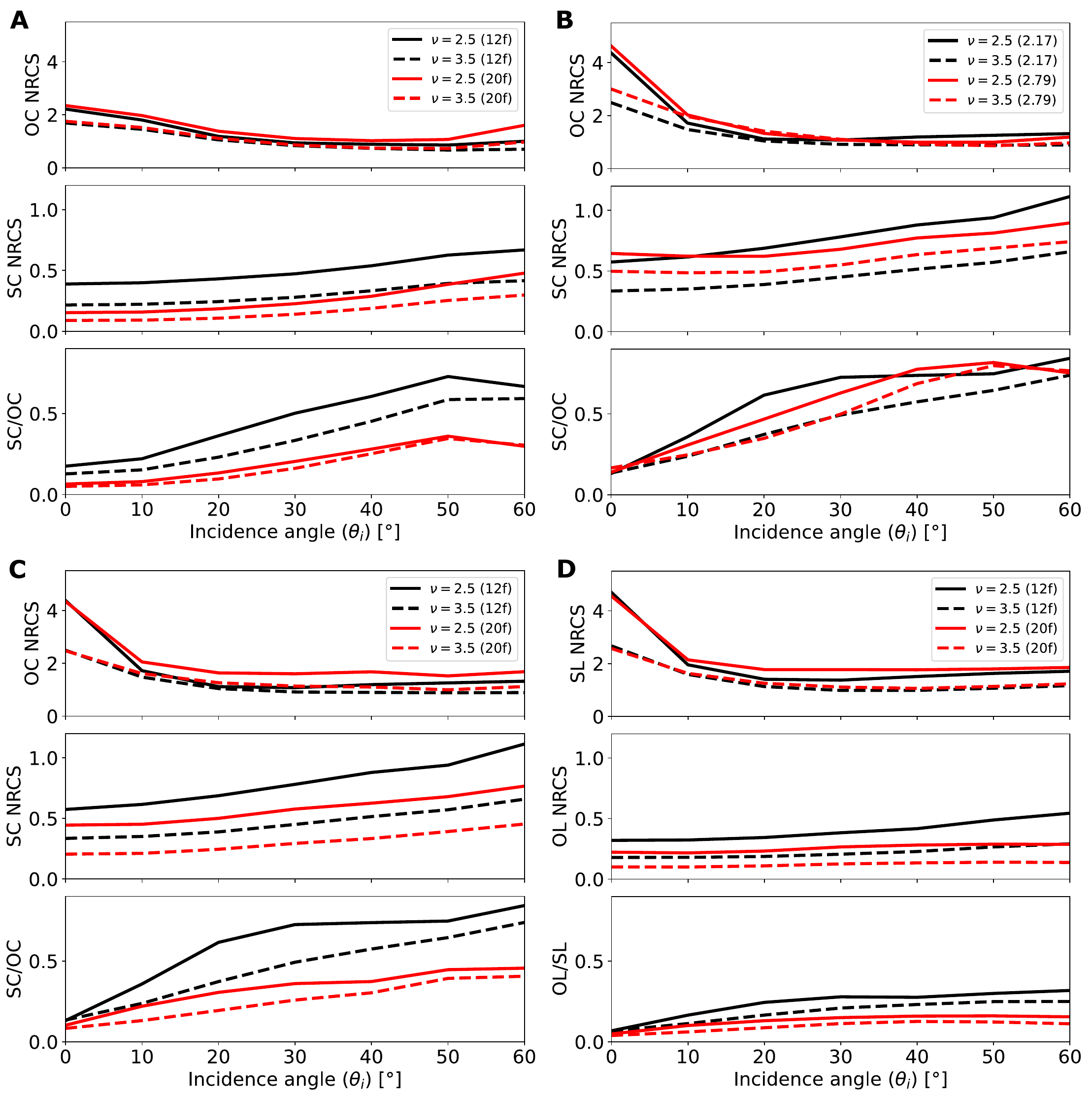}
	\caption{Sets A-C: The OC and SC NRCS and the SC/OC ratio for a power-law SFD of polyhedral particles for three different sets of parameters when the maximum size parameter is 3 (A) or 8 (B and C). The minimum size parameter is 0.5 in all cases. Set D: The same-linear (SL, on the top), orthogonal-linear (OL, in the middle) NRCS and the linear polarization ratio (LPR, on the bottom) for a power-law SFD with a size parameter range from 0.5 to 8. In A, C, and D, the number of faces is 12 (black) and 20 (red) while $m=2.17 + 0.004$i, whereas in B all lines are 12-face polyhedrons but the refractive index is either $2.17 + 0.004$i (black) or $2.79 + 0.0155$i (red). In all panels, the power-law index (for $x^{-\nu}$) is $\nu=2.5$ (solid lines) or $\nu=3.5$ (dashed lines)}
	\label{fig:Surfpoly_CP}
\end{figure*}

In this section, we focus more on the applications and present how the different physical properties affect the scattering properties when a size distribution is used instead of individual size parameters. Figures \ref{fig:Poly12vs20_SFD_pol} and \ref{fig:Poly217vs279_SFD_pol} show the scattering matrix elements for the two types of polyhedrons weighted using a power-law SFD $N(x) \propto x^{-3}$, where $x \in [0.5,8]$, while varying either the particle shape or refractive index. The use of SFD flattens the distinct forward-reflection diffraction peaks of the large size parameters, and stabilizes the polarization elements' oscillations of the small ones. Minor differences between the two polyhedral shapes remain, but are less evident than for the case of $x=3$ alone. 

The largest remaining differences are observed at back\-scattering directions, so we analyze them separately in Fig. \ref{fig:Surfpoly_CP}. The sets A-C show the OC and SC NRCS and the SC/OC ratio and the set D shows the corresponding linear polarization quantities for a power-law SFD of polyhedral particles when the maximum size parameter, the power-law index, the number of faces, and the refractive index are varied. The graphs show the roles that each physical parameter play in the observable quantities. The number of faces (particle roundness) plays a small role in the OC NRCS, but has a clear effect on the SC NRCS and the SC/OC ratio. 

Including the larger particles (up to $x=8$) or decreasing $\nu$ from 3.5 to 2.5 to increase the contribution of the larger particles has a visible effect on the OC NRCS at normal incidence, but otherwise mostly minor effect on all radar observables. The larger particles enhance all NRCSs, as was discussed in Section \ref{sec:sh_size}. In the SFD-weighted case, we see a generally increasing trend for SC and OL as a function of $\theta_i$, while for OC and SL the trend is decreasing or independent of $\theta_i$. The latter enhances the increasing trend of the SC/OC and OL/SL ratios.

Intriguingly, in the whole range of $\theta_i$, the SC/OC ratio seems only modestly affected by the choice of the maximum particle size and the power-law index, whereas the particle roundness plays a major role. Less rounded (12-face) polyhedrons also feature larger overall dependence of the SC/OC ratio on other parameters. The linear polarizations follow the same trends with only minor differences (Fig. \ref{fig:Surfpoly_CP}D): SL is positively correlated with OC, OL with SC, and $\mu_\mathrm{C}$ with $\mu_\mathrm{L}$, similar to ensemble and orientation-averaged particles in free space. The SL polarizations hh and vv are statistically equal at $\theta_i = 0^\circ$ and nearly equal with an accuracy of 85 \% at $\theta_i > 0^\circ$, whereas the OL polarizations hv and vh are equal with an accuracy of better then 95 \%.

For rough surfaces, the co-polarized ratio $\sigma_\mathrm{hh}/\sigma_\mathrm{vv}$ typically decreases as a function of incidence angle, while $\sigma_\mathrm{hv}$ and $\sigma_\mathrm{vh}$ are equal (see, e.g., \cite{ulaby2014}). Rough surfaces are discussed further in Section \ref{sec:discussion}.

\subsection{Significance of the surface}
\label{sec:surfacevsfree}

One of the key scopes of our investigation is the comparison of the backscattering values of the particles in free space and those on surfaces. In free space, the scattering properties are independent of the incidence angle, which is defined with respect to the normal of the surface when the latter is present. It is not a trivial question at which $\theta_i$, if at any, the backscattering properties of particles on a surface are close to those in free space.

One practical benefit of a potential agreement between the surface and free-space cases is the easier computations for the latter. While the simulations themselves are not that much slower in the surface case, thanks to efficient implementation in ADDA \citep{yurkin2015}, there are a number of other challenges introduced by the surface. A larger number of particle realizations is necessary as the surface constrains the angular space for orientation averaging (with current code capabilities), which requires more computational resources and hard-drive space. Also, the scattering quantities (cross sections and efficiencies) are not as universally defined, as we discussed in Section \ref{sec:theory}, which could complicate their use in applications and any general conclusions. Finally, the particle-substrate distance is an additional free parameter affecting the results (see Appendix \ref{sec:distance}). Using touching condition (as in this paper) allows us to determine this distance, but it requires an additional technical step and is not perfect with respect to realism of the resulting geometry.

\begin{figure}[ht]
	\centering
	\includegraphics[width=0.46\textwidth]{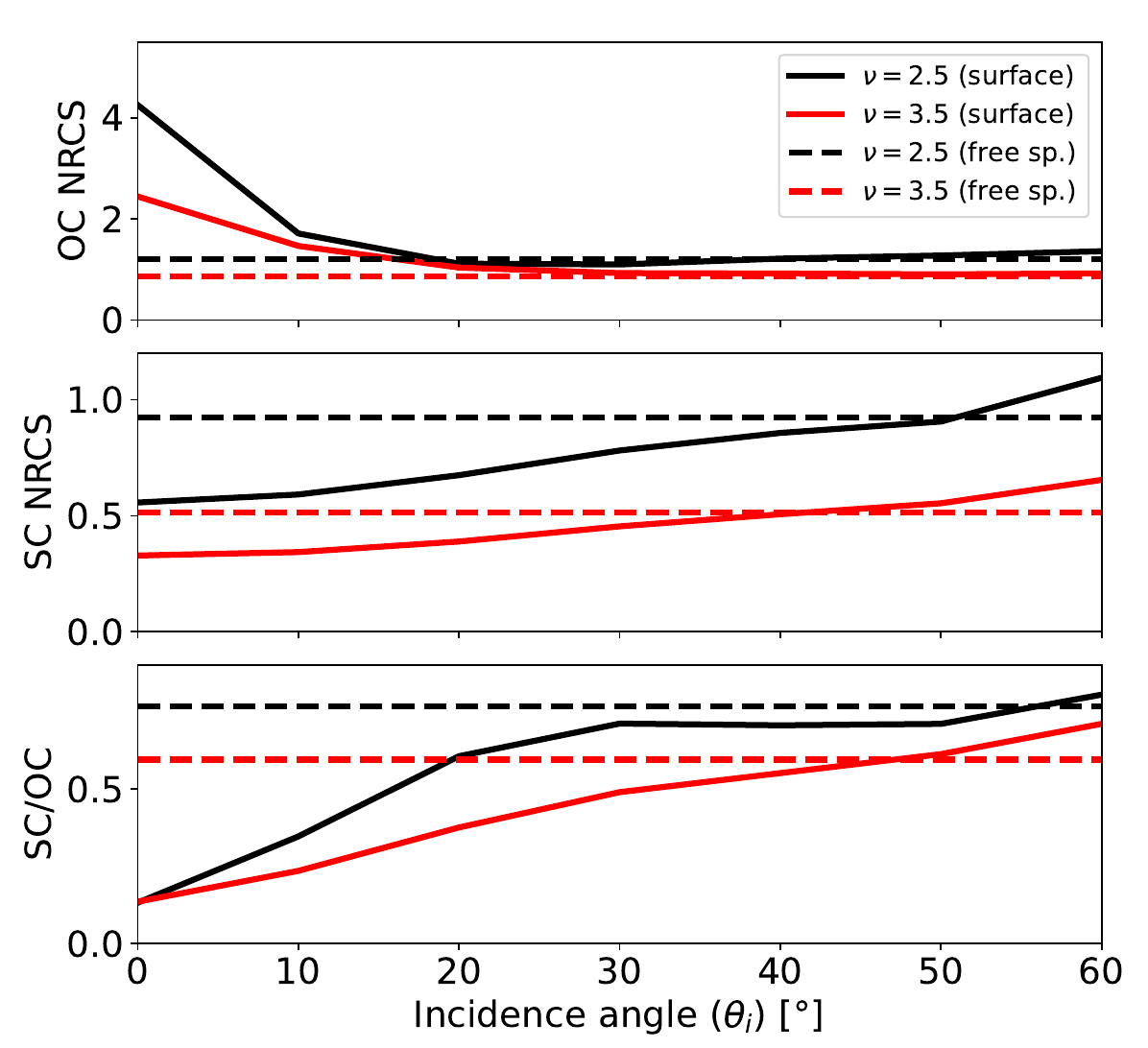}
	\caption{The power-law SFD-weighted OC NRCS, SC NRCS, and the SC/OC ratio of 12-face polyhedral particles in free space (dashed lines) and on a surface (solid lines) using $x \in [0.5,8]$ and $m=2.17+0.004$i. The value of the power-law index is labeled in the legend for each line color.}
	\label{fig:SurfVsFree}
\end{figure}

Figure \ref{fig:SurfVsFree} shows an example of the free-space case compared to the surface case using an SFD-weighted ensemble of 12-face polyhedrons with $m = 2.17 + 0.004$i. Scattering matrix elements of a few individual size parameters are shown as a function of scattering angle in Appendix \ref{sec:freespace}, while here we focus on the SFD-weighted results at backscattering. This result demonstrates that at incidence angles of $30^\circ$ to $50^\circ$ the polarization properties of particles on a surface are relatively close to those of particles in free space, whereas at $\theta_i < 20^\circ$, the difference is quite significant due to the reflected diffraction peak near normal incidence. At $\theta_i = 60^\circ$, the free-space results underestimate all the shown parameters.

We can also investigate the applicability of Eq. \ref{eq:cpr_lpr} for the surface case: we find it applicable to good accuracy (the relative difference between the left and right hand sides is $<15$\%) up to incidence angles of $40^\circ$, above which the accuracy depends on the particle shape and SFD. Counterintuitively, the 12-face polyhedrons with $\nu=2.5$ have $< 10$ \% error for Eq. \ref{eq:cpr_lpr} for all studied $\theta_i$, in contrast to the 20-face polyhedrons with $\nu=3.5$ having 44\% error at $\theta_i =60^\circ$, while the error of less than 10 \% only applies when $\theta_i \in [10^\circ,30^\circ]$. 

\begin{figure}[ht]
	\centering
	\includegraphics[width=0.46\textwidth]{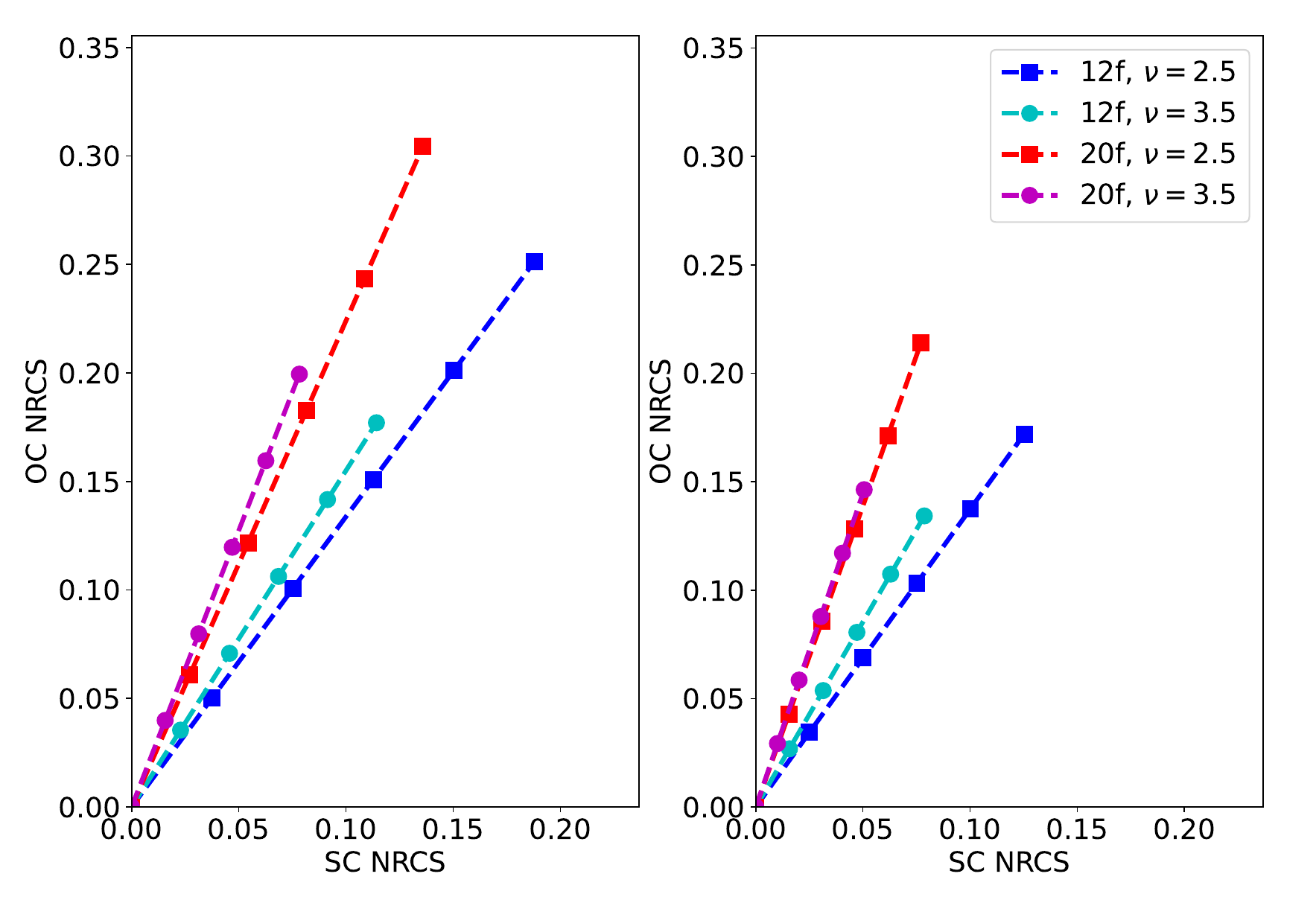}
	\caption{The OC NRCS versus the SC NRCS for a power-law SFD of polyhedral particles when the size parameter range is from 0.5 to 8 (on the left) or to 3 (on the right), the incidence angle is 50$^\circ$ and the surface-coverage ratio of the particles (as viewed from the normal) increases linearly from 0\% (bottom left markers) to 20\% (top right markers). The number of faces and the value of the power-law index are labeled in the legend for each line style.}
	\label{fig:Surfpoly_SCOC}
\end{figure}

Finally, in our last practical example case, we consider random ensembles with moderate surface coverage through simple superposition, weighted by their relative geometric cross section covering the surface (ignoring interaction between different particles). Specifically, Fig. \ref{fig:Surfpoly_SCOC} demonstrates an example of surfaces covered with different particle abundances (from 0\% to 20\% with five equal intervals) and two different ranges of particle sizes, illuminated at $\theta_i=50^\circ$. This approach was introduced for radar polarimetric analysis of lunar craters by \citet{virkki2019}, where scattering properties of compact but irregularly-shaped particles in free space were combined with that of the surface. In their study, the weighted total NRCS for any polarization (pol) for the unit surface  was estimated as $\hat{\sigma}_{\mathrm{tot,pol}} = P\hat{\sigma}_{\mathrm{par,pol}} + (1-P)\hat{\sigma}_{\mathrm{sub,pol}}$, where the subscripts sub and par refer to the substrate and particles, respectively, and $P$ is the surface-coverage ratio (the total geometric cross section $C_\mathrm{G}$ of polyhedrons per unit area of the surface). In this study, the surface is planar and the substrate is homogeneous, so $\hat{\sigma}_{\mathrm{sub,pol}}(\theta_i = 50^\circ) = 0$ for both polarizations. Despite somewhat different particles properties (both morphological and electric, with $\mathrm{Re}(m)=2.17$ versus $2.54$ in the cited paper), we find that Fig. \ref{fig:Surfpoly_SCOC} shows very similar trends to those in \citet{virkki2019}: a cone-shaped distribution of points, where the line slope depends (at a specific incidence angle) primarily on the particle properties such as the shape and the SFD, whereas the location along the line is primarily a function of the abundance of particles with specific properties. At $\theta_i = 50^\circ$, we find the slope varying from 1.34 for the 12-face polyhedrons with a power-law index $-2.5$ to 2.55 for the 20-face polyhedrons with a power-law index $-3.5$, when the size parameter range extends up to 8. The upper limit of the slope increases to 2.89 when the SFD range is limited to $x\leq3$, while the lower limit remains comparable at 1.37. 

Evidently, a more elaborate model will be required for interpreting observations due to a larger variety of scattering processes in most natural environments, as well as multiple-scattering effects for larger surface coverage. The weight coefficients used in the equation above omit the cross terms corresponding, e.g., to shadowing of the substrate by polyhedrons, and the surface area that contributes simultaneously to the backscattering by the surface and the particles through the pathways discussed in Section \ref{sec:theory}. The factor $P$ is independent of $\theta_i$ due to the used definition of $C_\mathrm{G}$ in Eq. \eqref{eq:sfd_weighted_nrcs} and the fact that we normalize by surface area instead of that of the incident beam. This contrasts with the apparent particle abundance, which obviously increases with $\theta_i$ and can be more relevant for another normalization of the experimental data. Furthermore, the increase of this apparent abundance should correlate with larger multi-particle effects.

\section{Discussion}
\label{sec:discussion}

To summarize the results for discussion, we have presented what roles the shape, size, and material of wavelength-scale particles on a surface play in their scattering properties, which has a variety of application from remote sensing to material physics. Similar to spheres in free space, spheres on a surface display strong resonance effects in contrast to non-spherical particles. The differences between 12-face and 20-face polyhedrons are more modest but clearly observable in the polarization properties of moderately small particles ($x\approx3$, Fig. \ref{fig:Poly_x6_12vs20_pol}), although it becomes negligible for the larger ones ($x\approx 6)$. Previous research has shown that, at least for refractive index up to 2.0, the particular shape plays a negligible role for ensembles of statistically similar wavelength-sized compact particles (e.g., \citep{muinonen2023}). For particle characteristics considered in this study, the statistical roundness of the particles plays a persistent role in the SFD-weighted polarization elements $F_{22}$ and $F_{44}$ at backscattering (Fig. \ref{fig:Poly12vs20_SFD_pol}).

Many of the symmetries expected for the random ensembles of particles in the free space are, to some extent, satisfied in the presence of the substrate. While this is highly dependent on the parameters of the problem, these symmetries provide some proxy to the relative strength of the particle-substrate interaction.  This is an interesting topic for future systematic studies. While we mostly discuss backscattering, there are also interesting conclusions for the specular direction. Specifically, for $\theta_i$ close to the Brewster's angle, the specular scattered light is almost perfectly polarized (Fig. \ref{fig:Poly_x6_12vs20_pol}) the same as for a smooth surface without the particle. This may be hypothetically used for estimating refractive index of a substrate sparsely covered with wavelength-scaled particles based on ellipsometry-type measurements with varying $\theta_i$ and $\theta_z$.



In terms of remote-sensing applications, the size-averaged scattering profiles provide the most useful results. We tested the roles of the size range (from $x=0.5$ up to $x=3$ versus up to $x=8$) and the power-law index (from $-2.5$ to $-3.5$) on the backscattering profiles for the two types of polyhedral particles. We found that the OC and SL enhancements near the normal incidence have the largest differences between the two size-parameter ranges. In the CPR, the effect of the size-parameter ranges (Fig. \ref{fig:Surfpoly_CP}A and C) is less than that of the power-law index, and both of them are less important than the shape. Thus, the particle shape plays the most pronounced role in our study. The material plays a small role as well (Fig. \ref{fig:Surfpoly_CP}B); however, the differences in OC and SC NRCS are within the computational uncertainties. Still, the role of permittivity can be significant when it is considered far outside the selected range (Fig. \ref{fig:Poly_x6_refri}).

The substrate was found to be a significant factor in the scattering profiles, which was not expected at high-incidence-angle backscattering. Nevertheless, for a power-law SFD-weighted case ($x^{-\nu}$, $\nu \in [2.5,3.5]$), the polarization elements were not greatly affected by the particle-substrate interactions at $\theta_i \in [30^\circ,50^\circ]$, and thus could also be approximated by ignoring the substrate to save computational resources. However, this is not a robust conclusion and can depend on parameters we did not investigate.

Because ADDA computes only the scattering properties of the particle, the scattering properties of the substrate must be considered separately if they are deemed significant for the total observables, as we showed in Fig. \ref{fig:Surfpoly_SCOC}. There is a clear benefit in the computational separation of the scattering properties of the particle and the surrounding substrate: the DDA is a versatile tool for single particles, but computationally very slow in the geometric-optics regime. Therefore, it is preferable to use other methods for the scattering properties of the substrate, which is likely to have surface roughness in many practical applications. By contrast, computational methods that explicitly consider (discretize) both the substrate (or its surface) and the particles typically suffer from diffraction effects from the limited surface width and have to limit the depth of the substrate, including previous attempts with the DDA \cite{parviainen2008,penttila2009}. 

For instance, the improved integral equation method for bidirectional scattering (IEM-B) in rough surfaces \citep{fung2002} provides an analytic approach with single-scale roughness (as opposed to fractal surfaces with multiscale roughness). Optionally, for semifractal surfaces (with scale limits to the applicability of specific fractal nature), a purely computational approach using Fresnel reflections from a triangularized topography model \cite{virkki2024} could be used to compute a far-field scattering matrix for the surface while assuming a semi-infinite substrate and thus avoiding unwanted diffraction effects. A computational code for IEM-B \citep{ulaby2014} has been made available in MATLAB\footnote{\url{https://mrs.eecs.umich.edu/microwave_remote_sensing_computer_codes.html}} and in Python\footnote{\url{https://github.com/ibaris/pyrism}}; however, neither code computes the scattering matrix directly but require some postprocessing. The Ray optics for self-affine fractal surfaces (ROSAS) code \cite{virkki2024} computes a scattering phase matrix for the vertical scattering plane by default but is limited to pure ray optics and requires ensemble averaging over a large number of topography realizations. 

Once both scattering matrices are computed, the total result would then be obtained as a weighted sum. The small fraction of the whole surface located under the particles, which interacts with the particle most strongly, can be assumed flat for simplicity (as in this paper), while the surrounding substrate may include modest roughness, realistic for natural surfaces. However, as discussed in Section \ref{sec:surfacevsfree}, the correct weights of the different scattering processes require careful consideration and thus, a significant amount of further research.

In all of the tested backscattering cases, we found the SC NRCS of the surface particles increasing as a function of $\theta_i$, while OC NRCS had a weaker dependency (when $\theta_i>10^\circ$). For example in lunar radar observations, the diffuse part of OC NRCS has been found to decrease as a function of $\cos^{1.5}(\theta_i)$, while the SC NRCS decreases as $\cos(\theta_i)$ (e.g., \cite{thompson2011}). Thus, the consideration of both particles and the substrate is required to model the observed behavior. The contribution of the surface particles may explain, in part, the difference in the incidence-angle dependency between the two polarizations.

Moreover, volume scattering from possible wavelength-scale scatterers or other permittivity variations below the surface would have to be added separately (not discussed in this paper). Developing a comprehensive simulation workflow accounting for all of the above factors, is a challenging open problem. For example, ensuring the conservation of energy will not be trivial, because the particle effectively interacts only with a part of the substrate and the different simulation domains (with and without particles) are not fully coupled. 

\section{Conclusions}
\label{sec:conclusions}
To conclude, ADDA's surface mode provides a rigorous computational approach to explore the scattering properties of a realistic rock shape model in a practically relevant scenario including the resonance-regime particles on a substrate. The differences between the scattering profiles of spherical and polyhedral particles, especially near backscattering, confirms the well-known fact that realistic particle shapes are critical for the simulation of polarimetric properties of wavelength-scale particles. 

We found only minor differences between the two primarily tested materials (refractive indices $2.17+0.004$i and $2.79 + 0.0155$i or, respectively, permittivities $4.7+0.016$i and $7.8+0.09$i), as well as indications of larger differences when the refractive index is much lower. Thus, the role of the permittivity would require more systematic research, e.g., for typical values at optical wavelengths. The particle shape was found to have a small but observable effect on the polarimetric properties near backscattering even for an SFD-weighted case, so that statistically more angular polyhedrons with less faces have larger CPR and LPR. Also, power-law index $-2.5$ elevated all observable circular-polarization components as well as CPR and LPR in comparison to $-3.5$. Limiting the size range to $x=3$ (in contrast to $x=8$) decreased the OC and SC NRCS but did not have a systematic effect on CPR. Therefore, CPR is not as good diagnostic parameter to particle SFD as the NRCSs. 

We investigated whether the surface is needed in the computations, or whether simulations of particles in free space can approximate the polarization properties at backscattering at moderate-to-high incidence angles (thus, outside the specular reflection direction). Although the polyhedral particles on a surface and in free space have similar backscattering for $\theta_i$ from $30^\circ$ to $50^\circ$, this agreement depends heavily on $\theta_i$ and was tested only for the 12-face polyhedrons. The scattering-matrix symmetries known to apply to ensemble- and orientation-averaged particles were applicable to the surface particles at smaller $\theta_i$ (up to about $40^\circ$), but we found larger discrepancies in the symmetries especially for the rounder particles at larger $\theta_i$. Therefore, if accurate scattering simulations of particles on surfaces are needed, the rigorous method presented in this paper is strongly recommended. 

We were only able to scratch the surface of the large parameter space, including the physical properties (shape, roughness, permittivity) for both the particles and the substrate, as well as their mutual contributions to observed scattering through the number density and size-frequency distribution of the particles covering the surface. Therefore, we leave an open door for future research for many other applications that may benefit from the same approach.

\section*{Data and code availability}
The ensemble-averaged and azimuthally averaged scattering matrices are available through Zenodo at \url{https://doi.org/10.5281/zenodo.15040282}. The codes for reading and visualizing the data published in Zenodo are available in GitHub at \url{https://github.com/a-virkki/ADDA-grid-plot-codes}. The ADDA code is available in GitHub at \url{https://github.com/adda-team/adda}. We have updated it with an example to compute the scattering properties of a spherical particle on a surface and to reproduce Figure 8 and parts of Figure 11 of this paper, available at \url{https://github.com/adda-team/adda/tree/master/examples/papers/2025_surface}. 

\section*{Acknowledgments}
AV acknowledges funding by the Research Council of Finland grant No. 347627. MY acknowledges support of the Normandy Region (project RADDAERO).

\section*{Declaration of competing interest}
The authors declare the following financial interests/personal relationships which may be considered as potential competing interests: Maxim Yurkin is an associate editor of JQSRT.
\section*{Appendix}
\appendix
\section{Accuracy tests}
\label{sec:accuracy}

In this section, we show in more detail what role the choice of the number of dipoles per wavelength (dpl) for the discretization and the number of realizations in the ensemble averages plays in the scattering properties, and thus, the presented results. 

\begin{figure}[ht]
  \centering
	\includegraphics[width=0.45\textwidth]{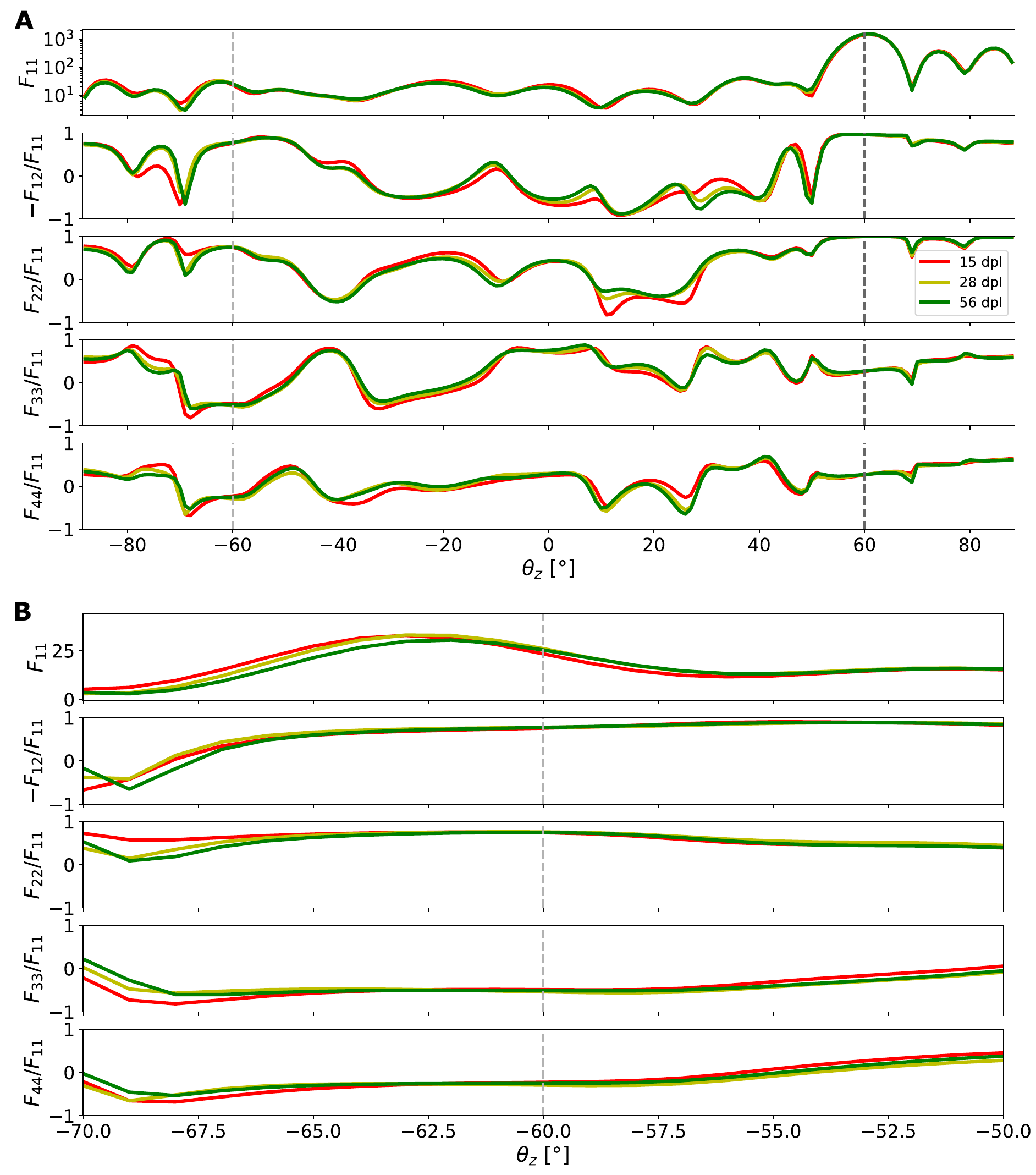}
	\caption{The scattering matrix elements $F_{11}$, $-F_{12}/F_{11}$, $F_{22}/F_{11}$, $F_{33}/F_{11}$, and $F_{44}/F_{11}$ in the vertical scattering plane as a function of zenith angle for 12-face polyhedrons with $m = 2.17 + 0.004$i, when $x=9$ and the incidence angle is 60$^\circ$. The dpl value varies from 15 (red) to 28 (yellow) and further to 56 (green). The panels above show all zenith angles within $89^\circ$ from zero, while the panels below are focused within $10^\circ$ from the backscattering direction at $\theta_z = -60^\circ$.}
	\label{fig:DplTest}
\end{figure}

Fig. \ref{fig:DplTest} shows the scattering matrix elements for one individual 12-face polyhedron with $x=9$ and $m = 2.17 + 0.004$i, when the number of dipole discretization is varied from 15 dpl (below the recommended value of 22) to 28 dpl and further to 56 dpl. Although the differences are minor between all cases at backscattering, there are clear discrepancies between 15 and 28 dpl, especially at high zenith angles, and minor discrepancies between 28 and 56 dpl. Based on this test, we conclude that 15 dpl would be too rough a discretization level. The results presented in the paper have been computed using more than 50 dpl for the most contributing particles with $x<4$ and at minimum 21 dpl when $m = 2.17 + 0.004$i, and can thus be considered reliable, i.e. unlikely to have differed if a greater dpl had been used.

\begin{figure}[htb]
  \centering
	\includegraphics[width=0.45\textwidth]{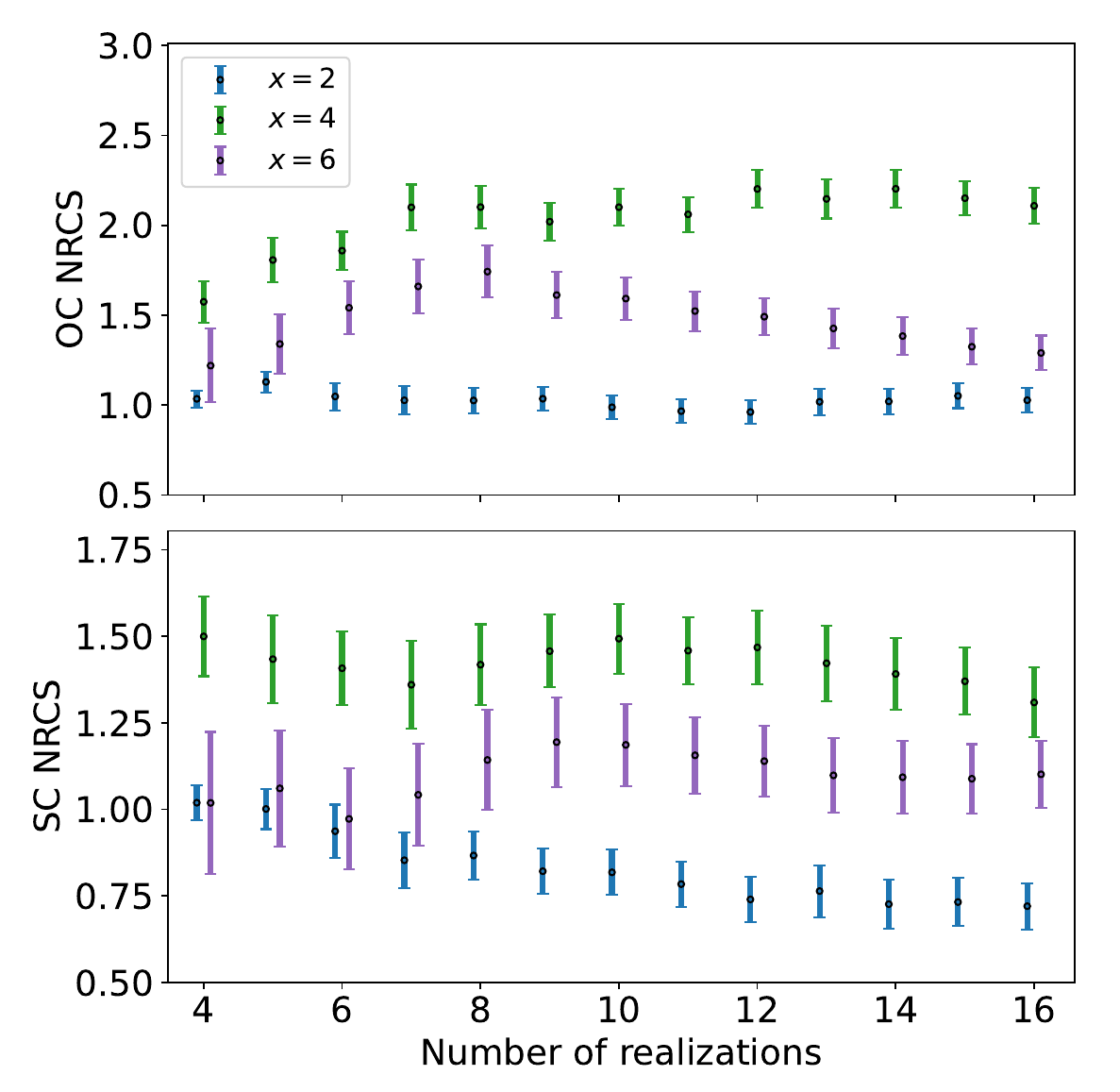}
	\caption{The uncertainties of the OC and SC NRCS for 12-face polyhedrons with $m = 2.17 + 0.004$i and $x=2$ (blue, with a horizontal offset of $-0.1$), $x=4$ (green), and $x=6$ (purple, with a horizontal offset of $+0.1$), as functions of the number of realizations. The incidence angle is 50$^\circ$. The black circles and the error bars depict respectively the means and the standard error of the mean for azimuthally and ensemble-averaged polarization elements.}
	\label{fig:EnsTest}
\end{figure}

\begin{figure}[htb]
  \centering
	\includegraphics[width=0.45\textwidth]{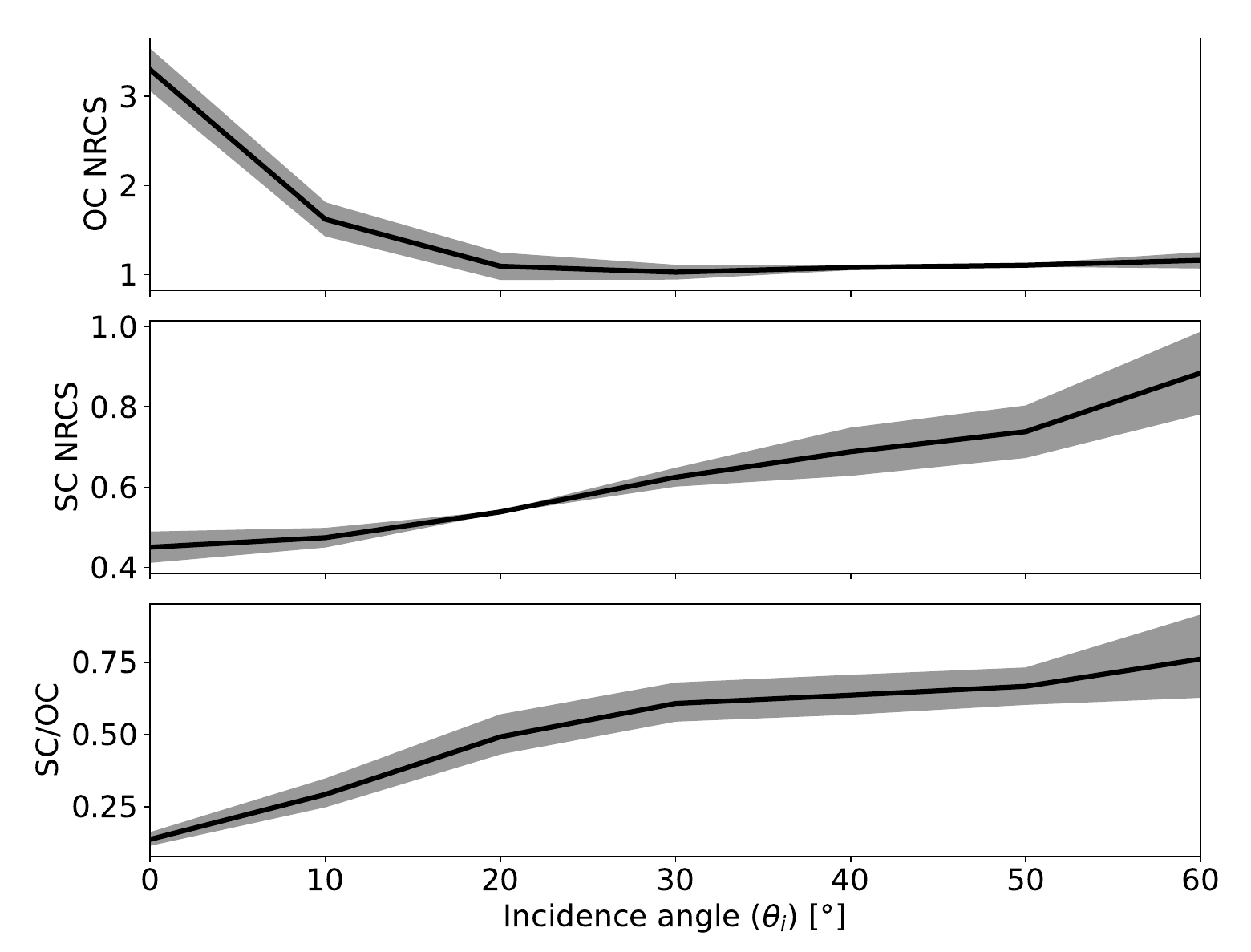}
	\caption{The uncertainties of the OC and SC NRCS for 12-face polyhedrons with $m = 2.17 + 0.004$i for azimuthally and ensemble-averaged, SFD-weighted scattering matrices over a range of incidence angles similar to Fig. \ref{fig:Surfpoly_CP}B-C, but using a power-law SFD $N(x) \propto x^{-3}$. The gray regions show the limits of the average curves for two independent sets of eight realizations each, while the black curve depicts the ensemble average of all 16 realizations.}
	\label{fig:EnsTest_SFD}
\end{figure}

Figures \ref{fig:EnsTest} and \ref{fig:EnsTest_SFD} illustrate how the number of realizations used for the ensemble averages affects the precision of the results. Despite the azimuthal average applied to all shown values, the number of realizations has a clear impact on the standard error of the mean values of both the opposite (OC) and same-circular (SC) particle-size-normalized radar cross sections (NRCS). These error bars were obtained by the propagation of standard deviations of the azimuthally averaged scattering-matrix elements and division by the square root of the number of realizations. In the OC polarization, the values change distinctively as a function of the number of realizations, which demonstrates the importance of using ensemble averages of more than a dozen particles for the 12-face polyhedrons. The standard error of the mean is greater for larger size parameters as the shape variations and features become more pronounced with respect to the wavelength. However, the power-law weighting reduces the relative role of the larger size parameters, including the corresponding uncertainties, as Fig. \ref{fig:EnsTest_SFD} demonstrates. Here, two separate sets of eight realizations were used to illustrate the possible uncertainties in an SFD-weighted case. Preliminary tests showed negligible covariance for the azimuthally averaged scattering matrix elements $F_{11}$ and $F_{44}$. Much larger number of realizations could improve the accuracy of results, but only slightly so. Note that the required hard drive space for the presented data set is already over 40~GB.

\section{Surface roughness}
\label{sec:roughness}
\begin{figure*}[htb]
  \centering
\includegraphics[width=0.9\textwidth]{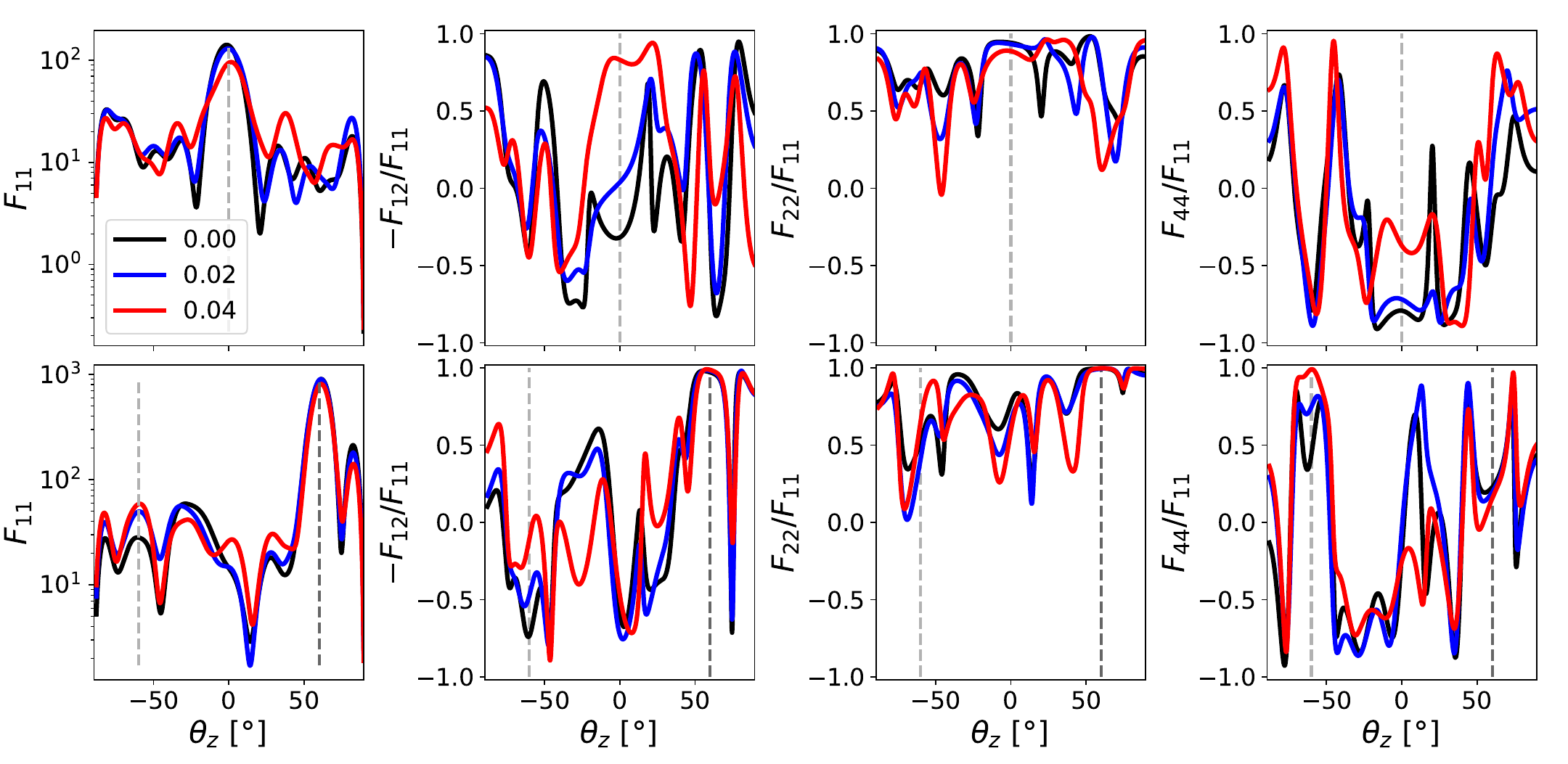}
	\caption{The scattering matrix elements $F_{11}$, $-F_{12}/F_{11}$, $F_{22}/F_{11}$, and $F_{44}/F_{11}$ in the vertical scattering plane as a function of zenith angle for 12-face polyhedrons with $m = 2.17 + 0.004$i ($x=8$) when using different levels of surface roughness: flat faces (0; black), modest roughness (0.02 used throughout the paper; blue), and very rough (0.04; red). The incidence angle is 0$^\circ$ (the top row), and 60$^\circ$ (the bottom row), both in the same realization in a fixed orientation, for which only the roughness changes. The vertical light gray and dark gray dashed lines depict, respectively, the backscattering and specular directions (coinciding for $\theta_i = 0^\circ)$.  See Section \ref{sec:methods} for more details on the definition of the roughness.}
	\label{fig:roughness}
\end{figure*}

The effect of surface roughness on the (single-orientation) scattering matrix elements of one particle for two incidence angles is shown in Fig. \ref{fig:roughness}. There are evident differences due to the surface roughness, although we use here only one particle realization in a fixed orientation, which hampers interpretation. The literature shows that the role of roughness is especially noticeable in the $F_{22}$ and $F_{44}$ elements at backscattering, with enhanced contribution from larger particles \cite{kemppinen15}. Therefore, the effect of particle surface roughness on the results should not be fully ruled out. 

In the case of the 60$^\circ$ incidence, one of the faces of the particle is oriented at the incidence direction; yet, no clear backscattering enhancement is observed for the case of flat faces compared to the rougher ones. By contrast, at backscattering, the greatest $F_{11}$ is that of the roughest particle.

\section{Particles in free space}
\label{sec:freespace}

\begin{figure}[htb]
  \centering
\includegraphics[width=0.45\textwidth]{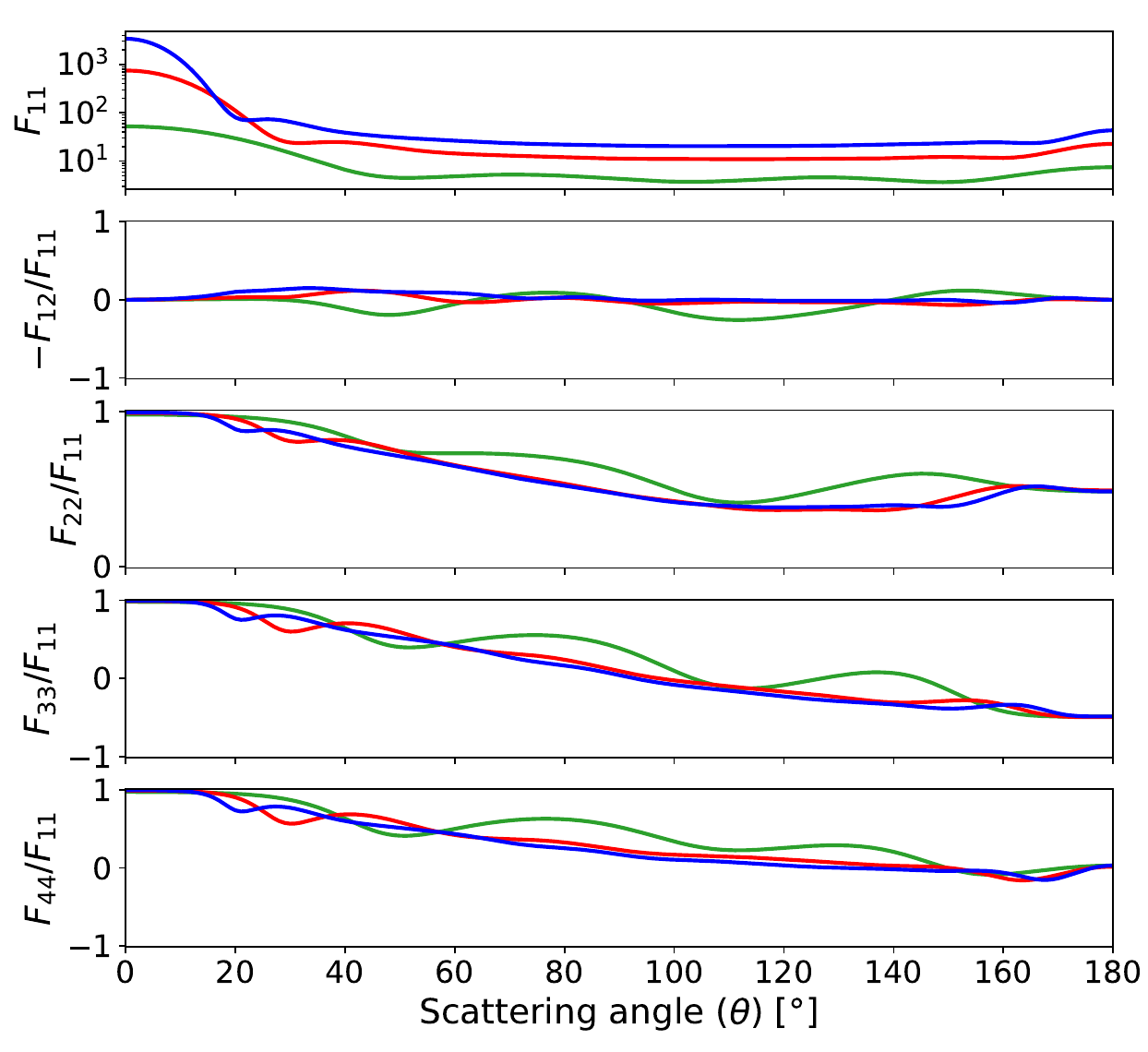}
	\caption{The scattering matrix elements $F_{11}$, $-F_{12}/F_{11}$, $F_{22}/F_{11}$, $F_{33}/F_{11}$, and $F_{44}/F_{11}$ as functions of scattering angle (with respect to the forward-scattering direction) for an ensemble of twelve 12-face polyhedrons in free space averaged over 81 $\times$ 32 orientations, when $m = 2.17 + 0.004$i and $x=3$ (green), 6 (red), and 9 (blue).}
	\label{fig:poly12_freespace}
\end{figure}

In this section, we provide additional information of a free-space reference case by using an ensemble of twelve 12-face polyhedrons with $m = 2.17 + 0.004$i and a few selected size parameters (Figure \ref{fig:poly12_freespace}). Here, the scattering matrix is computed with respect to the forward-scattering direction using orientation averaging for 81 $\times$ 32 orientations (as described in Section \ref{sec:methods}). The scattering matrix has a symmetric, block-diagonal form through all angles. The corresponding diagonal elements correlate with each other (for each size) and, when divided by $F_{11}$, are independent of the particle size at backscattering (when $x \in [3,9]$). Moreover, they satisfy backscattering symmetry rules \citep{hovenier2000}:  $F_{22}=-F_{33}$ and $F_{44} = F_{11} - 2F_{22}$.  Also, the similarity of the polarization elements of the cases $x=6$ and $x=9$ is notable when $\theta>60^\circ$. This can be useful for efficient simulation of SFD-averaged properties in free space, since the large size parameters are increasingly time-consuming. 

\section{Particle distance from the surface}
\label{sec:distance}
Here we show how the distance $d$ of the particle center from the surface affects the observed scattering properties. This particular scenario may not seem directly relevant for the key scope of this paper; however, it is interesting in terms of electromagnetic scattering in general, since it allows us to disentangle far-field and near-field interaction of the particle with the substrate (Section \ref{sec:theory}). The latter must decay with increasing $d$.

The scattering matrices were computed for a 12-face polyhedron in a fixed orientation with $x=8$ and $m=2.17 + 0.004$i at varying multiples of wavelength from the underlying substrate: 0 (touching), $\lambda/3$, $\lambda$, and $3\lambda$.  We also consider the same particle in free space with the same direction of incidence (and definition of the scattering angles), so that the only difference between the results for each $\theta_i$ ($0^\circ$ or $60^\circ$) is where the substrate is located (if present at all).

\begin{figure*}[htb]
  \centering
	\includegraphics[width=0.95\textwidth]{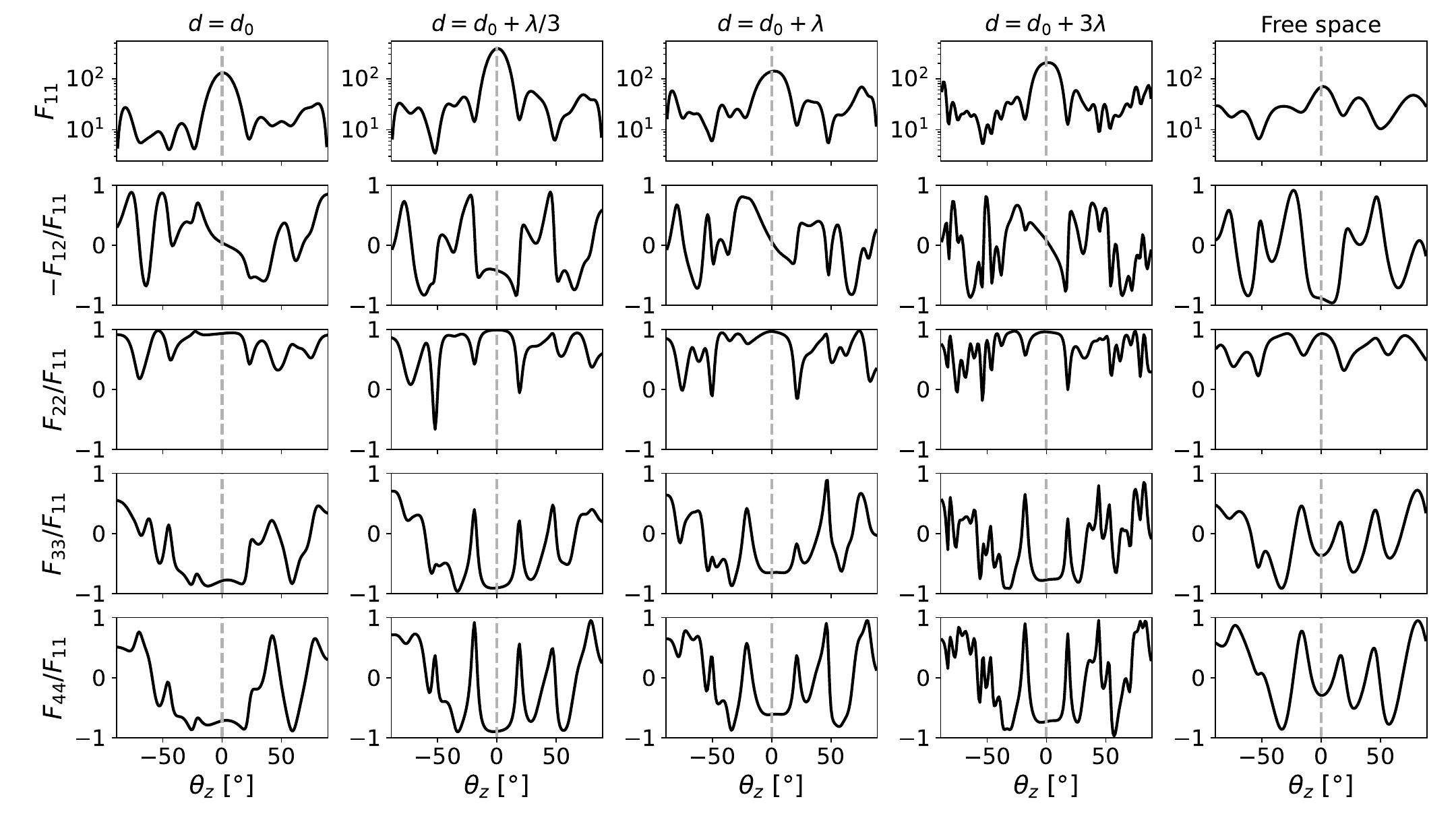}
	\caption{The scattering matrix elements $F_{11}$, $-F_{12}/F_{11}$, $F_{22}/F_{11}$, $F_{33}/F_{11}$, and $F_{44}/F_{11}$ in the vertical scattering plane as a function of zenith angle for a 12-face polyhedron in fixed orientation with $m = 2.17 + 0.004$i, when $x=8$ and the incidence angle is 0$^\circ$. The distance of the particle center from the surface grows as annotated at the top of columns 1-4, while the fifth column is for a particle in free space. At distance $d_0$, the particle touches the surface.}
	\label{fig:DistTest_i0}
\end{figure*}

\begin{figure*}[htb]
  \centering	\includegraphics[width=0.95\textwidth]{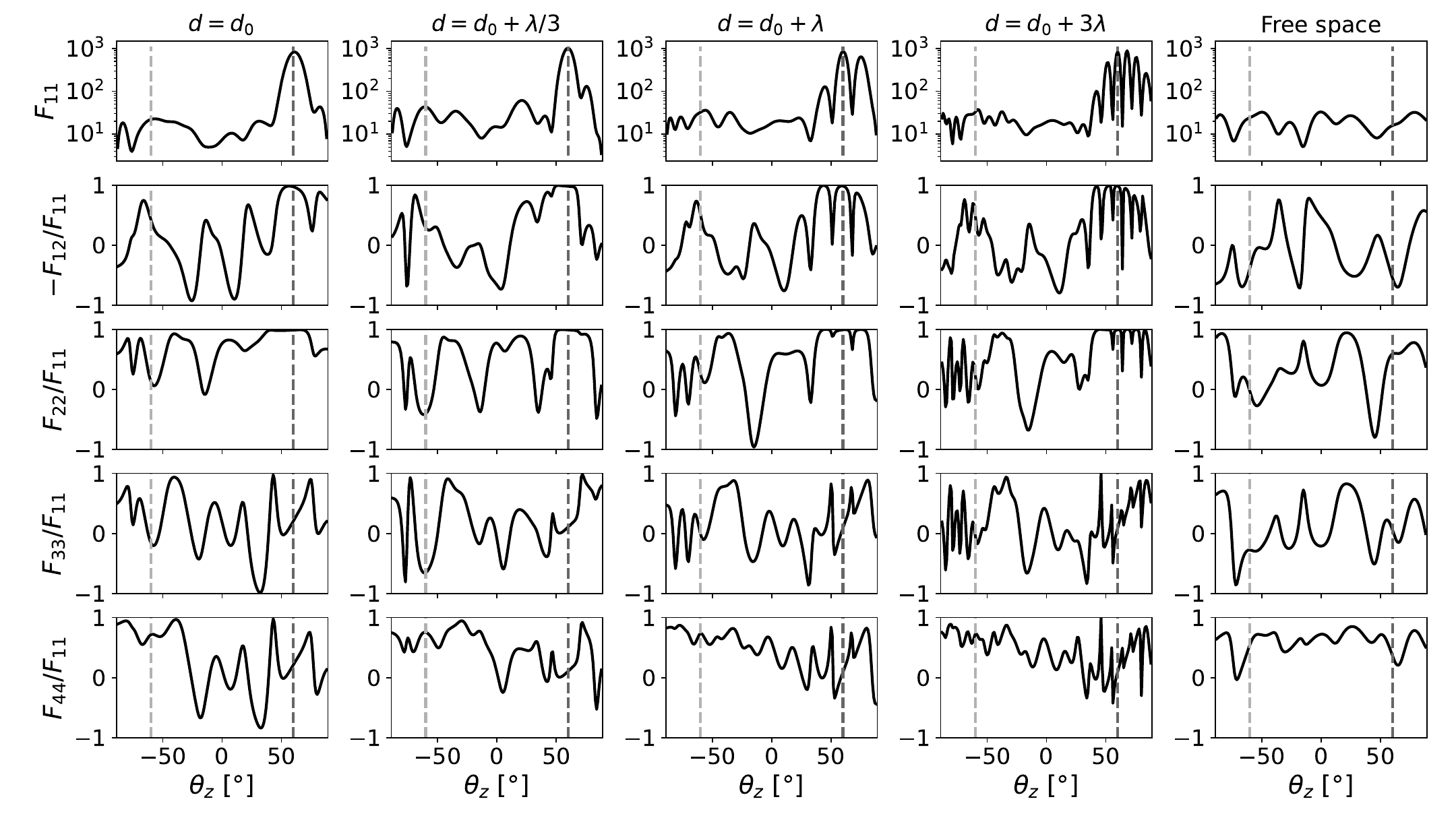}  
	\caption{The same as Fig. \ref{fig:DistTest_i0}, but using an incidence angle of 60$^\circ$.}
	\label{fig:DistTest_i60}
\end{figure*}

Figures \ref{fig:DistTest_i0} and \ref{fig:DistTest_i60} show how the distance $d$ affects the frequency of the oscillations as a function of $\theta_z$ for the selected scattering matrix elements. This far-field interference effectively results from the presence of the particle and its mirror image. Thus, the period of oscillations is determined by the projection of the corresponding distance on the plane perpendicular to the scattering direction, specifically, it is approximately equal to $\lambda/{[2d\sin(|\theta_z|)]}$. The fringe contrast is proportional to the relative intensity of the mirror image and, thus, to the reflection coefficients for both incident and scattered radiation. Therefore, the contrast increases with both $\theta_i$ and $|\theta_z|$. These trends are supported by the figures, where the most vivid illustration is the modulation of the specular peak for $\theta_i = 60^\circ$.

Comparing the results for different distances (columns 1--4 in Figs. \ref{fig:DistTest_i0}
and \ref{fig:DistTest_i60}), we see that near-field particle-substrate interaction has little systematic effect on the results. At least, it is smaller than the far-field interference increasing with $d$. The latter can potentially be decreased by averaging over a range of distances, as was suggested for direct discretization of the substrate \cite{penttila2009}, but that is outside the scope of this paper. Still, in terms of locations of large-scale minima and maxima, the free-space result is closer to that for $d=d_0+3\lambda$ (as expected). However, the comparison for $\theta_i = 0^\circ$ and small $\theta_z$ is hampered by overlap of backscattering and specular reflection, while for $\theta_i = 60^\circ$, the free-space simulation obviously does not reproduce the specular reflection peak. By contrast, the agreement between the free-space and $d=d_0+3\lambda$ results is the best for $\theta_i = 60^\circ$ and $|\theta_z| < 30^\circ$.







\printcredits  

\bibliographystyle{elsarticle-num-names}

\bibliography{refs}



\end{document}